\documentstyle[prb,aps]{revtex}
\begin{document}
\draft
\tighten
\title{Analytical results on quantum interference and magnetoconductance 
for strongly localized electrons in a magnetic field: exact summation of 
forward-scattering paths }

\author{Yeong-Lieh Lin}
\address{Department of Physics, The University of Michigan, 
Ann Arbor, Michigan 48109-1120 \\
and Department of Physics, West Virginia University, 
Morgantown, West Virginia 26506-6315\,\cite{lyl}}
\author{Franco Nori}
\address{Department of Physics, The University of Michigan, 
Ann Arbor, Michigan 48109-1120\,\cite{lyl} \\
and Institute for Theoretical Physics, University of California, 
Santa Barbara, California 93106-4030}
\date{\today}
\maketitle
\begin{abstract}
We study quantum interference effects 
on the transition strength
for strongly localized electrons 
hopping on two-dimensional (2D) square and three-dimensional (3D) 
cubic lattices in the presence of a magnetic field. 
These effects arise from the interference between phase
factors associated with different electron paths connecting two distinct 
sites. For electrons confined on a square lattice, with and without disorder, 
we obtain {\em closed-form} expressions for the tunneling probability, 
which determines the conductivity, 
between two {\em arbitrary} sites by exactly summing the
corresponding phase factors of {\em all} forward-scattering paths connecting
them. By analytically 
summing paths which allow backward 
excursions in the forward-scattering direction, we find that the 
interference patterns between the dominant winding paths are not 
drastically different from those between the directed paths. 
An {\em analytic field-dependent} expression, valid in {\em any dimension}, 
for the magnetoconductance (MC) is derived. A  {\em positive} MC is clearly 
observed when turning on the magnetic field. 
In 2D, when the strength of ${\bf B}$ reaches a certain value, which is 
inversely proportional to twice the hopping length, 
the MC is increased by a factor of two compared to that at 
zero field. The periodicity in the flux of the MC is found to be 
equal to the superconducting flux quantum $hc/2e$.
We also investigate transport on the much less-studied and 
experimentally important 3D cubic lattice case, 
where it is shown how the interference patterns and the small-field 
behavior of the MC vary according to the orientation of the applied field 
${\bf B}$. At very small fields, for two 
sites diagonally separated  a distance $r$, we find that the MC behaves as: 
$rB$ in quasi-1D systems, $r^{3/2}B$ in 2D with ${\bf B}=(0,0,B)$, and 
$rB$ [$r^{3/2}B$] in 3D with ${\bf B}$ parallel [perpendicular] 
to the $(1,1,1)$ direction. Furthermore, for a 3D sample, the effect on the 
low-flux MC due to  the randomness of the angles between the 
hopping direction and the orientation of ${\bf B}$ is examined analytically.
\end{abstract}
\widetext

\section{Introduction}
Electrons moving on a lattice immersed in a
magnetic field have attracted much attention due to their relevance to 
many physical problems.  In particular, quantum interference (QI) effects 
between different electron paths in disordered electron systems  have been 
a subject of intense study 
because they play an important role in quantum transport. For instance, 
the QI of closed loops and their time-reversed paths is  central to 
weak-localization phenomena.\cite{weak} Indeed, 
during the past decade and half, 
many fascinating  phenomena---including universal conductance fluctuations 
as well as magnetic-field and spin-orbit scattering 
effects on the conductivity---observed 
in the {\em weakly} localized, metallic regime have been understood in 
terms of the 
QI between different Feynman diffusive paths in backscattering loops 
(i.e., paths bringing an electron back to the starting point). 
Recently, interest 
has grown in the effects of 
a magnetic field on {\it strongly} localized 
electrons\cite{1,2,3,4,5,6,7,rmt} with 
variable-range hopping (VRH) where striking QI phenomena has been observed 
in mesoscopic and macroscopic
insulators or strongly disordered compounds: 
anomalous magnetoresistance, pronounced conductance fluctuations,  
Aharonov-Bohm oscillations with periods of $hc/e$ and $hc/2e$,
and the Hall effect. 
This strongly localized regime\cite{1,2,3,4,5,6,7,rmt} is less 
well-understood than the weak-localization case.

In the strongly localized regime, the major mechanism for transport 
is thermally activated hopping between the localized sites. 
In the VRH introduced by Mott,\cite{mott} localized electrons, 
whose wavefunctions decay exponentially with a localization 
length $\xi$, hop a distance which is 
many times larger than $\xi$. As a result of the balance between the 
probabilities for hopping and thermal activation, Mott derived that 
in $d$ dimensions the hopping length changes with temperature as 
$\xi(T_0/T)^{1/(d+1)}$, where $T_0$ is a characteristic temperature. 
Therefore, the lower the temperature is, the further away the electron 
tunnels in order to  find a localized site of closer energy.

According to the ``critical path analysis"\cite{8} arguments, 
the conductance of the sample is governed by one critical hopping event. 
During this critical phonon-assisted tunneling process, 
the electron traverses many other 
impurities since the hopping length is very large at low 
temperatures. While encountering these intermediate scatterers, 
the electron preserves its phase memory. This elastic multiple-scattering 
is the origin of the QI effects associated with a single hopping event 
between the initial ($i$) and final ($f$) sites. 
The overall tunneling amplitude $T_{if}$ between 
the sites $i$ and $f$  is therefore the sum of the contributions from 
all possible paths connecting them.\cite{1,2} In other words, the tunneling 
probability of one distant hop is determined by the interference of 
many electron paths between $i$ and $f$.  This leads to 
Mott's law for the temperature dependence of the conductivity 
in $d$ dimensions:\cite{mott}
$$\sigma(T) \sim |T_{if}|^{2}\,\exp
\left[-\left(\frac{T_0}{T}\right)^{\frac{1}{d+1}}\right].$$
It is worthwhile to note that, in the limit of strong localization, 
the dominant contribution to $T_{if}$ comes from the shortest paths 
between $i$ and $f$ (i.e., the ``directed path model"). In other words, only 
interference between {\em forward-scattering} paths 
need to be taken into account. 
This is in  contrast with weak localization which results from 
{\em backscattering} processes on closed paths. The focus of this paper 
is on the QI effects on $T_{if}$ and relevant physical quantities due 
to the presence of an external magnetic field. 
We will use the model proposed in Ref.~3, 
which is used in  most of the recent 
theoretical work in this area. In this model, 
the impurities are arranged on a regular 
square (cubic) lattice in 2D (3D) and a nearest-neighbor tight-binding 
Anderson Hamiltonian is employed.

In this work we investigate the QI of strongly localized electrons 
by doing {\em exact} summations over {\em all} 
forward-scattering paths between two {\em arbitrary} sites. 
We derive compact {\em closed-form} expressions for various physical 
quantities (e.g., the transition strength which determines the conductivity)
for electrons propagating on a square lattice 
subject to an external magnetic field, with and without random impurities. 
We also obtain an explicit formula for an experimentally important case that 
has been much less studied theoretically so far: the interference between 
paths on 
a 3D cubic lattice.
 
In the disordered case, by analytically computing the moments for the 
tunneling probability and employing the replica method, we derive 
{\em analytic} results for the magnetoconductance (MC) 
in terms of {\em sums-over-paths}, which are 
applicable in {\em any dimension}. Our explicit field-dependent expressions 
for the MC provide a precise description of the MC in terms of the 
magnetic flux. A {\em positive} MC, with a saturation value slightly 
larger than  twice the MC at zero field, is observed when turning on 
the field ${\bf B}$.
In 2D, the saturated value of the 
magnetic field $B_{{\rm sa}}$ (i.e., the first field that 
makes the MC become twice the value at zero field) is 
inversely proportional to twice the hopping length: the larger the system is, 
the smaller $B_{{\rm sa}}$ will be. In other words, as soon as the 
system, with hopping distance $r$, is penetrated by a total flux 
of $(r/8)(hc/e)$, the MC reaches the saturation value. 
The period of oscillation of 
the MC is found to be equal to $hc/2e$, which is the superconducting 
flux quantum. 
 
Furthermore, at very small fields, 
for two sites diagonally separated  a distance $r$, 
the MC scales as follows: 
(i) $r\,B$ for quasi-1D ladder-type geometries with ${\bf B}=(0,0,B)$, 
(ii) $r^{3/2}\,B$ in 2D with ${\bf B}=(0,0,B)$, 
(iii) $r\,B$  in 3D with ${\bf B}$ parallel to the $(1,1,1)$ direction, 
and (iv) $r^{3/2}\,B$ in 3D with ${\bf B}$ perpendicular to the $(1,1,1)$ 
direction.

The general expressions presented here:
(i) {\it contain, as particular cases}, several QI 
results\cite{1,2,3,4,5,6,7}  
derived during the past decade 
(often by using either numerical or approximate methods),
(ii) include QI to arbitrary points $(m,n)$ on a square lattice, 
instead of only diagonal sites $(m,m)$, 
(iii) focus on 2D {\em and} 3D lattices, 
and (iv) can be extended to also include {\em backward} excursions 
(e.g., side windings) in the directed paths.

Exact results in this class of directed-paths problems are valuable and can 
be useful when studying other systems, for instance: (1) directed polymers in 
a disordered substrate (see, e.g., Refs.~12 and 13), (2) interfaces in 
2D (see, for instance, Ref.~14), (3) light propagation in random 
media,\cite{dashen,feng} and (4) charged bosons in 1D.\cite{blum}

To study the magnetic field effects on the tunneling probability 
of strongly localized electrons, we start from the tight-binding Hamiltonian
\begin{equation}
H=W \sum_{i} c_{i}^{\dag}c_{i} + V \sum_{\langle ij \rangle}
c_{i}^{\dag}c_{j} \exp(i A_{ij})\,,
\end{equation}
where $\langle ij \rangle$ refers to the nearest-neighbor sites and the phase 
$A_{ij}=2\pi \int_{i}^{j}{\bf A}{\cdot}d{\bf l}$ is $2\pi$ times the line 
integral of the vector potential along the bond from $i$ to $j$ in units of 
the normal flux quantum $hc/e$. 
In the strongly localized regime, $V/W \ll 1$, 
the electron energy can be set to zero.\cite{3,5} 
Consider two states, localized at sites $i$ and $f$ which 
are $r$ bonds apart.  By using a locator 
expansion, the transition amplitude (Green's function) 
$T_{if}$ between these two states can be expressed as\cite{1,2,3,4,5,6,7} 
\begin{equation}
T_{if}=\sum_{l=0}^{\infty}\,W\,\left(\frac{V}{W}\right)^{r+2l}\,S^{(r+2l)}\,,
\end{equation}
where
\begin{equation}
S^{(r+2l)}= \sum_{\stackrel{\scriptstyle {\rm All}\ (r+2l)
\rm{-step\ paths}\ \Gamma}
{\scriptstyle\rm{connecting\ sites}\ \em{i}\ \rm{and}\ \em{f}}} 
\ e^{i\Phi_\Gamma}\,,
\end{equation}
and $\Phi_{\Gamma}$ is the sum over phases of the bonds on the path 
$\Gamma$ of $r+2l$ steps connecting sites $i$ and $f$. In general, 
$\Gamma$ contains closed loops. In the strongly localized regime 
(i.e., $V/W \ll 1$), the dominant 
contribution to $T_{if}$ is $W(V/W)^{r}S^{(r)}$, where
\begin{equation}
S^{(r)}= \sum_{\stackrel{\scriptstyle\rm{All\ directed\ paths\ \Gamma}}
{\scriptstyle\rm{of}\ \em{r}\ \rm{steps\ on\ a\ lattice}}} 
\ e^{i\Phi_\Gamma}\,,
\end{equation}
In other words, only the shortest-length paths 
(with {\em no} backward excursions) connecting them 
are taken into account, namely, the 
{\em directed-path model} of Refs.~2-8. This directed-path model provides an 
excellent approximation to $T_{if}$ since $(V/W)^{2}$ is quite small in the 
extremely localized regime.\cite{1,2,3,4,5,6,7} 
It is important to stress that the {\em conductivity}\cite{1,2,3,4} 
between $i$ and $f$ is proportional to $|T_{if}|^{2}$. 

Quantum interference, contained in 
$S^{(r+2l)}$, arises because the phase factors of different paths 
connecting the initial and final sites interfere with each other. 
We will first focus on the computation of $S^{(r)}$, which is the essential 
QI quantity for electrons deep in the localized regime. In 2D, we also 
analytically compute $S^{(r+2)}$ which becomes important 
when electrons are not so strongly localized.

This paper is organized as follows. 
In Sec.~II, we study QI on a square 
lattice under a uniform potential, which is related to the decay of gap 
states into the bulk.\cite{1} Here we derive an elegant, general, and very 
compact closed-form expression 
for $S^{(r)}$. Intriguing properties associated with the behavior of 
$S^{(r)}$ on diagonal sites are discussed in detail. 
It will be shown later (in section~III) that the effect of a magnetic field on 
the MC is governed by the behavior of $S^{(r)}$. 
As a step towards the understanding of interference between 
non-directed  paths, 
we also go beyond the directed-path model by exactly computing 
analytic results for $S^{(r+2)}$.

In Sec.~III, we investigate the tunneling in a random impurity potential, 
which is relevant to the conductivity of, for example, 
lightly  doped semiconductors and strongly disordered compounds.\cite{1} 
Closed-form results for the tunneling probability, which determines the 
conductivity, are obtained. We then analytically compute 
the moments for the tunneling probability. From them, we derive analytic 
field-dependent expressions, valid in any dimension, for 
the MC. The full behavior of the MC as a function of the magnetic 
flux---including the scaling in the low-field limit and 
the occurrence of saturation--- is discussed in detail. The close relationship 
between the QI quantity $S^{(r)}$ and the corresponding MC is illustrated. 
Comparison of our results with experimental observation and other 
theoretical work is also made.

In Sec.~IV, we examine the QI on a 3D cubic lattice and provide a 
general formula for $S^{(r)}$. 
We show how the interference patterns and the small-field 
behavior of the MC vary according to the orientation of the applied field. 
Furthermore, we investigate the effect on the low-flux MC due to 
the randomness the angles between the directions of the 
critical hops  and the orientation of the applied field.
 
In Sec.~V, we conclude by addressing several relevant issues and summarize 
our results.

\section{Quantum Interference on a two-dimensional 
square lattice}
\subsection{Exact summation of forward-scattering paths: $S^{(r)}$}
Let $(m,n)$ denote the site coordinates.  Without 
loss of generality, we choose $(0,0)$ to
be the initial site and focus on $m,n \geq 0$. For forward-scattering
paths of $r$ steps, which exclude backward excursions (i.e., only
moving upward and to the right is allowed), ending sites $(m,n)$
satisfy $m+n=r$. Let $S_{m,n}$ (=$S^{(r)}$) be the sum over all directed
paths of $r$ steps on which an electron can hop from the origin to
the site $(m,n)$, each one weighted by its corresponding phase factor.
Employing the symmetric gauge 
$${\bf A}=\frac{B}{2}(-y,x),$$ 
and denoting the flux through an elementary plaquette 
(i.e., with an  area corresponding to the square of the average distance, 
which is typically equal to or larger than the localization length $\xi$, 
between two impurities) by $\phi/2\pi$, 
it is straightforward to construct the recursion relation: 
\begin{equation}
S_{m,n}=e^{-in\phi/2}\,S_{m-1,n}\,+\,e^{im\phi/2}\,S_{m,n-1}\,. 
\end{equation}
This equation states that the site $(m,n)$ can be reached by taking the 
$r$th step from neighboring sites to the left or below. The factors 
in front of the $S$'s, namely, $\exp(-in\phi/2)$  and $\exp(im\phi/2)$, 
account for the presence of the magnetic field. 
Enumerating the recursion relations for $S_{k_{n},n}$ 
($k_{n}=m-1,m-2,\ldots,0$) successively and using $S_{0,n}=1$ 
for any $n$, we obtain the following relation 
\begin{equation}
S_{m,n}=\sum_{k_n=0}^{m}\,e^{ik_n\phi/2}\,
  e^{-i(m-k_n)n\phi/2}\,S_{k_n,n-1}\,.
\end{equation}
Here $S_{m,n}$ is expressed as a sum of the $S$'s one row below 
(i.e., on the line $y=n-1$). The physical meaning of Eq.~(6) 
is clear: the site $(m,n)$ can be reached by moving one step upward 
from sites $(k_{n},n-1)$ with $0 \leq k_{n}\leq m$, acquiring the 
phase $ik_{n}\phi/2$; then traversing $m-k_n$ steps from $(k_n,n)$ 
to $(m,n)$, each step with a phase $-in\phi/2$. 
By applying Eq.~(6) 
recursively (and utilizing $S_{m,0}=1$ for 
any $m$), $S_{m,n}$ for $m,n \geq 1$ can be written as
\begin{equation}
S_{m,n}(\phi)=e^{-i mn \phi/2}\,L_{m,n}(\phi)\,,  
\end{equation}
where
\begin{eqnarray}
L_{m,n}(\phi)&=&\sum_{k_n=0}^{m}
\sum_{k_{n-1}=0}^{k_{n}}\cdots\sum_{k_1=0}^{k_2}
\,e^{i(k_1+\cdots+k_{n-1}+k_{n})\phi} \\  \nonumber
&=&\prod^{n}_{j=1}\,\sum^{k_{j+1}}_{k_j=0}\,e^{ik_j\phi}\,,
\end{eqnarray}
with $k_{n+1}\equiv m$. If we use the Landau gauge instead, the expression 
for the sum-over-paths $S^{(r)}$ will read $L_{m,n}$; namely, 
$S_{m,n}$ employs the {\em S}ymmetric gauge, 
while $L_{m,n}$ uses the {\em L}andau gauge. Notice that 
each term in the summand corresponds to the overall phase 
factor associated with a directed path. In the absence of the 
magnetic flux ($\phi =0$),
\begin{eqnarray}
S_{m,n}(0)&=& \sum_{k_n=0}^{m}\sum_{k_{n-1}=0}^{k_{n}}\cdots
\sum_{k_1=0}^{k_2}\,1 \nonumber \\
&=& C_{m}^{m+n}=\frac{(m+n)!}{m!\,n!}\equiv N,
\end{eqnarray}
which is just the total number of $r$-step paths between $(0,0)$ 
and $(m,n)$.

After some calculations we obtain one of our main results, 
a very compact and elegant closed-form expression for $S_{m,n}(\phi)$:
\begin{equation}
S_{m,n}(\phi)=\frac{F_{m+n}(\phi)}{F_{m}(\phi)\,F_{n}(\phi)}\,,
\end{equation}
where
\begin{equation}
F_{m}(\phi)=\prod_{k=1}^{m}\sin\frac{k}{2}\phi\,.
\end{equation}
Notice that the symmetry $S_{m,n}=S_{n,m}$ [apparent in Eq.~(10)] is due
to the square lattice geometry. Moreover, we also obtain 
\begin{equation}
L_{m,n}(\phi)=\frac{\prod_{k=1}^{m+n}\left(1-e^{ik\phi}\right)}
{\left[\,\prod_{k=1}^{m}\left(1-e^{ik\phi}\right)\right]
\left[\,\prod_{k=1}^{n}\left(1-e^{ik\phi}\right)\right]}\,.
\end{equation}

Previous work on QI in the VRH regime obtained 
particular cases, mostly numerical, of sums to diagonal points $S_{m,m}$, 
while the general result Eq.~(10) is valid for arbitrary 
(e.g., non-diagonal) sites.  

To illustrate the quantum interference originating from sums over 
phase factors associated with directed paths, we show the five possible 
ending sites $(m,n)$ for $r=4$ and their corresponding $S_{m,n}$ in Fig.~1(a). 
In addition, the six different paths connecting $(0,0)$ and
$(2,2)$ and their separate phase factor contributions to $S_{2,2}$ are
shown in Fig.~1(b). 

\subsection{Low-flux limit}
In the very-low-flux limit $\phi \ll 1$, the logarithm of
$S_{m,n}$, calculated exactly to order $\phi^2$, is
\begin{equation}
\ln S_{m,n}(\phi) = \ln N-\frac{1}{24}m\,n\,(m+n+1)\,\phi^{2}\,.
\end{equation}
Thus we obtain the familiar\cite{1}  harmonic 
shrinkage of the wave function with {\em explicit} expressions 
for all the prefactors. This result can be interpreted as follows. The 
effective ``cigar-shape" area exposed to the field has an effective 
length $l_{{\rm eff}}$: 
$$l_{{\rm eff}}\sim \sqrt{mn}$$ 
(i.e., the square root of the area enclosed by the paths) and an 
effective width $w_{{\rm eff}}$: 
$$w_{{\rm eff}}\sim \sqrt{m+n}$$ (i.e., the square root of the 
length of the paths). 
For the special case $m=n=r/2$, 
\begin{equation} 
\ln S_{m,m}(\phi) = 
\ln \frac{r!}{[(r/2)!]^{2}}-\frac{1}{96}r^{2}(r+1)\,\phi^{2}\,,
\end{equation}
which is consistent with, and generalizes, the results 
in Ref.~4 since it 
gives the exact prefactor. Thus, the effective length is $\sim r$, while 
the effective width is $\sim\sqrt{r}$. Furthermore, for a ladder-type quasi-1D 
system, (e.g., $m=r-1$ and $n=1$), we have
\begin{equation} 
\ln S_{r-1,1}(\phi) = 
\ln r-\frac{1}{24}(r^{2}-1)\phi^{2}\,.
\end{equation}
In this case, the effective length and width are both $\sim \sqrt{r}$. 
This result remains  valid for small values of $n$ (narrow stripes 
or multiladders).
The fourth-order contribution to $\ln S_{m,n}(\phi)$ 
can also be computed exactly as
\begin{eqnarray*}
&\frac{m\,n}{103680}\,[\,10\,(6m^4+15m^3n+20m^2n^2+15mn^3+6n^4)& \\
&+114\,(m^3+2m^2n+2mn^2+n^3)& \\
&+29\,(2m^2+3mn+2n^2)+9\,(m+n)+5\,]\,\phi^4.&
\end{eqnarray*}

\subsection{Quantum interference on diagonal sites}
Among the $S$'s for an even number of steps, those located along the
diagonal corners contain the richest interference effects since the
number of paths ending at $(m,m)$ and the areas they enclose are both the 
largest. We therefore examine
more closely the behavior of the quantities
\begin{equation}
I_{2m}(\phi)\ \equiv\ S_{m,m}(\phi)\ =\ \prod_{k=1}^{m}\,
\frac{\sin\frac{m+k}{2}\,\phi}{\sin\frac{k}{2}\,\phi}\,.
\end{equation}
For irrational flux $\phi$, it can be proved that $-1 < I_{2m} < 1$ 
for any $m$. A {\em particular case} (asymptotic behavior) of our very 
compact general expression Eq.~(16) for $I_{2m}$ is investigated in detail 
by Fishman, Shapir, and Wang in Ref.~6. 
\ For $\phi=2\pi\,s/t$ ($s$ and $t$ are positive integers 
with $1 \leq s <t$ and $s$ being prime to $t$),
we obtain ($n \geq 0$) for $m < t$ 
\begin{equation}
I_{2(m+nt)}=\frac{(-1)^{stn}(2n)!}{n!\,n!}\,I_{2m}\,,
\end{equation} and
\begin{equation}
I_{2nt}=(-1)^{stn}\frac{(2n)!}{(n!)^2}\,.
\end{equation} 
Furthermore, for those $m$ satisfying 
\begin{equation}
\frac{t}{2} \leq m \leq t-1\,,
\end{equation}
\begin{equation}
I_{2(m+nt)}=0\,.
\end{equation} In other words, the zeros of $I_{2m}(\phi)$ are given
by $\phi=2\pi\,s/t$ for 
\begin{equation}
\frac{m+1}{n+1} \leq t \leq \frac{2m}{2n+1}\,,
\end{equation}
with
\begin{equation}
0 \leq n \leq \frac{m-1}{2}\,,
\end{equation} 
and the $s$'s are prime to each allowed
$t$. From a physical viewpoint, these flux values produce the complete 
cancelation of all phase factors (i.e., {\em fully destructive interference}) 
and result in the {\em vanishing} of the tunneling probability 
(and conductivity). 
Indeed, we will see in section~III that if we also consider the effects 
of the on-site impurity scattering, these flux values lead to the 
largest (i.e., saturated) value for the positive MC.

In Fig.~2, we show the zeros for $I_{2}, I_{4}, I_{6}, \dots, I_{40}$ obtained 
by using Eqs.~(21) and (22). Note that the smallest value of $\phi/2\pi$ 
satisfying $I_{2m}(\phi)=0$ is always $1/2m$ and the 
number of zeros rapidly increases when $m$ becomes larger.

The $I_{2m}$ can be expressed as sums of trigonometric cosines. For instance, 
the first few ones are (with $\theta \equiv \phi/2$):
\begin{eqnarray*}
I_{2}&=&2\cos\theta, \\
I_{4}&=&2+2\cos2\theta+2\cos4\theta, \\
I_{6}&=&6\cos\theta+6\cos3\theta+4\cos5\theta+2\cos7\theta
+2\cos9\theta, \\
I_{8}&=&8+14\cos2\theta+14\cos4\theta+10\cos6\theta+10\cos8\theta
+6\cos10\theta+4\cos12\theta+2\cos14\theta+2\cos16\theta, \\  
I_{10}&=&40\cos\theta+38\cos3\theta+36\cos5\theta+32\cos7\theta
+28\cos9\theta+22\cos11\theta+18\cos13\theta \\
& &\mbox{}+14\cos15\theta+10\cos17\theta+6\cos19\theta+4\cos21\theta
+2\cos23\theta+2\cos25\theta, \\
I_{12}&=&58+110\cos2\theta+110\cos4\theta+102\cos6\theta
+96\cos8\theta+84\cos10\theta+78\cos12\theta+64\cos14\theta \\
& &\mbox{}+56\cos16\theta+44\cos18\theta+36\cos20\theta+26\cos22\theta
+22\cos24\theta+14\cos26\theta \\
& &\mbox{}+10\cos28\theta+6\cos30\theta+4\cos32\theta
+2\cos34\theta+2\cos36\theta, \\
I_{14}&=&338\cos\theta+332\cos3\theta+324\cos5\theta+310\cos7\theta
+292\cos9\theta+272\cos11\theta+250\cos13\theta+224\cos15\theta  \\
& &\mbox{}+200\cos17\theta+174\cos19\theta+150\cos21\theta+126\cos23\theta
+106\cos25\theta+84\cos27\theta+68\cos29\theta+52\cos31\theta  \\
& &\mbox{}+40\cos33\theta+30\cos35\theta+22\cos37\theta+14\cos39\theta
+10\cos41\theta+6\cos43\theta+4\cos45\theta+2\cos47\theta+2\cos49\theta.
\end{eqnarray*}
Notice that $I_{2m}$ depends only on the even (odd) harmonics of
$\theta$ when $m$ is even (odd). $I_{2m}(\phi)$ obeys the following properties:
(i) $2\pi$ ($4\pi$) periodicity in $\phi$ for even (odd) $m$, namely, 
\begin{eqnarray*}
I_{4n}(\phi+2\pi)&=&I_{4n}(\phi)\,, \\
I_{4n+2}(\phi+4\pi)&=&I_{4n+2}(\phi)\,.
\end{eqnarray*}
In other words, the period of $I_{2m}$ corresponds to $hc/e$ when 
$m$ is even, and $2hc/e$ when $m$ is odd. 
(ii) With $m$ even
$$I_{2m}(2\pi-\phi)=I_{2m}(\phi)$$ 
for $0 \leq \phi \leq \pi$. Also $$I_{2m}(\pi)=\frac{m!}{[(m/2)!]^2}.$$ 
(iii) With $m$ odd
$$I_{2m}(2\pi\pm\phi)=-I_{2m}(\phi)$$ 
for $0 \leq \phi \leq 2\pi$.

From the properties described above, we can draw a general 
picture of the behavior of $I_{2m}$. Let $\Phi \equiv \phi/2\pi$. 
$I_{2m}(\Phi=0)=(2m)!/(m!)^{2}$, 
which is a enormous number for large $m$. 
As the magnetic field is turned on, $I_{2m}$ rapidly drops to 
its first zero at $\Phi=1/2m$. $I_{2m}$ then shows distinct 
behaviors depending on $m$.

For even $m$, $I_{2m}$ exhibits many small-magnitude fluctuations around 
zero for $\frac{1}{2m} < \Phi < \frac{m-1}{2m}$. $I_{2m}$ then monotonically 
climbs from $0$ to a large positive value, $I_{2m}(\pi)=m!/[(m/2)!]^2$, for 
$\frac{m-1}{2m} \leq \Phi \leq \frac{1}{2}$. It is evident that 
$I_{2m}(\pi)$ is still very small compared to $I_{2m}(0)$. 
Within the period $0 \leq \Phi \leq 1$ (i.e., $0 \leq \phi \leq 2\pi$), 
$I_{2m}$ has mirror symmetry with respect to $\Phi=1/2$ (i.e., $\phi=\pi$).

For odd $m$, $I_{2m}$ exhibits many small-magnitude 
fluctuations around zero for $\frac{1}{2m} < \Phi < \frac{2m-1}{2m}$. 
In addition, $I_{2m}$ always equals $0$ at $\Phi=1/2$ (i.e., $\phi=\pi$). 
$I_{2m}$ then monotonically drops from $0$ to $-(2m)!/(m!)^{2}$ for 
$\frac{2m-1}{2m} \leq \Phi \leq 1$. For $0 \leq \Phi \leq 1$, $I_{2m}$ has 
inversion symmetry with respect to $\Phi=1/2$. 
Within the period $0 \leq \Phi \leq 2$ (i.e., $0 \leq \phi \leq 4\pi$), 
$I_{2m}$ has mirror symmetry around $\Phi=1$ (i.e., $\phi=2\pi$). Recall 
that for any $m$, $-1 < I_{2m} < 1$, for irrational values of $\Phi$.

In Fig.~3, we plot 
$I_{2}$ through $I_{12}$, $I_{18}$, $I_{20}$, $I_{38}$ and $I_{40}$. 
These figures show very interesting interference patterns of $I_{2m}$ 
and clearly reflect the general description given above. It is worthwhile 
to keep in mind that the properties embedded  in $S_{m,n}$ 
described above play a central role in determining the behavior of the MC 
obtained in section~III.

\subsection{Exact summation of the dominant winding paths: $S^{(r+2)}$}
Up to now, we have focused on the computation of $S^{(r)}$ and presented a 
detailed investigation of their properties. When the electrons are less 
strongly localized, the next higher-order contribution to $T_{if}$ 
(i.e., $W(V/W)^{r+2}S^{(r+2)}$, which is the dominant 
term {\em including backward excursions}) becomes important. 
Therefore, quantum interference effects 
between phase factors of paths with backward recursions 
(i.e., moving downward and to the left is also included) need to be taken 
into consideration. Notice that paths in $S^{(r+2)}$, though include 
backscattering processes, do not involve closed loops enclosing flux 
(e.g., elementary square plaquettes).

In this section we present the computation of the second-order contribution, 
namely $S^{(r+2)}$, to the transition amplitude $T_{if}$. Let $P_{m,n}$ 
($=S^{(r+2)}$) denote the sums over paths of $m+n+2$ steps starting from 
$(0,0)$ and ending at $(m,n)$. We assume that the electrons are confined on 
a square lattice with non-negative $x$ and $y$ coordinates. We can divide 
the contribution to $P_{m,n}$ into five parts.

First, hopping directly to site 
$(p,0)$, with $1 \leq p \leq m$, 
electrons take one step back to $(p-1,0)$, then hop 
$m-p+1+n$ steps to $(m,n)$. Second, hopping directly to site $(0,q)$, 
with $1 \leq q \leq n$, 
electrons take one step back to $(0,q-1)$, then hop 
$m+n-q+1$ steps to $(m,n)$. Third, directly hopping to 
site $(p,n+1)$, with $0 \leq p \leq m$, 
electrons move one-step downward to $(p,n)$ gaining a phase factor 
$\exp(-ip\phi/2)$, 
then hop $m-p$ steps to $(m,n)$. Fourth, directly hopping to 
site $(m+1,q)$, with $0 \leq q \leq n$, 
electrons move one-step downward to $(m,q)$ gaining a phase factor 
$\exp(iq\phi/2)$, then hop $n-q$ steps to $(m,n)$.  Fifth, 
directly hopping to $(p,q)$ 
with $1\leq p \leq m$ and $1\leq q \leq n$, electrons take one step back to 
either $(p-1,q)$ or $(p,q-1)$, accompanied by the phase factor 
$\exp(iq\phi/2)$ or $\exp(-ip\phi/2)$, then hop $m+n-p-q+1$ steps 
to the ending site $(m,n)$.
Therefore $P_{m,n}$ can be written as
\begin{eqnarray}
P_{m,n}&=&\sum_{p=1}^{m}B_{p-1,0\rightarrow m,n}
+\sum_{q=1}^{n}B_{0,q-1\rightarrow m,n}
+\sum_{p=0}^{m}S_{p,n+1}\,e^{-ip\phi/2}\,B_{p,n\rightarrow m,n}
+\sum_{q=0}^{n}S_{m+1,q}\,e^{iq\phi/2}\,B_{m,q\rightarrow m,n} \nonumber \\
& &\mbox{}+\sum_{p=1}^{m}\sum_{q=1}^{n}S_{p,q}\left(
e^{iq\phi/2}\,B_{p-1,q\rightarrow m,n}
+e^{-ip\phi/2}\,B_{p,q-1\rightarrow m,n}\right). 
\end{eqnarray}
where $B_{p,q\rightarrow m,n}$ is the sum over phase factors of all directed 
paths (i.e., containing $m+n-p-q$ steps) starting from $(p,q)$ 
and ending at $(m,n)$. After some calculation we obtain
\begin{equation}
B_{p,q\rightarrow m,n}=\exp\left\{i\frac{[-(m-p)q+(n-q)p]\phi}{2}\right\}
\,S_{m-p,n-q}.
\end{equation}
By substituting Eq.~(24) into Eq.~(23), we derive
\begin{eqnarray}
P_{m,n}&=&\sum_{p=1}^{m}e^{\frac{in(p-1)\phi}{2}}S_{m-p+1,n}
+\sum_{q=1}^{n}e^{-\frac{im(q-1)\phi}{2}}S_{m,n-q+1}
+e^{-\frac{imn\phi}{2}}\sum_{p=0}^{m}e^{\frac{i(n-1)p\phi}{2}}S_{p,n+1}
+e^{\frac{imn\phi}{2}}\sum_{q=0}^{n}e^{-\frac{i(m-1)q\phi}{2}}S_{m+1,q} 
\nonumber \\
& &\mbox{}+\sum_{p=1}^{m}\sum_{q=1}^{n}\,S_{p,q}\left\{
e^{-i[(m-1)q-n(p-1)]\phi/2}\,S_{m-p+1,n-q}
+e^{-i[m(q-1)-(n-1)p]\phi/2}\,S_{m-p,n-q+1}\right\}. 
\end{eqnarray}
In the special case $m=n$,
\begin{eqnarray}
P_{m,m}&=&2\,\sum_{j=1}^{m}\cos\left[\frac{m(m-j)\phi}{2}\right]\,S_{j,m}
+2\,\sum_{j=0}^{m}\cos\left[\frac{(m^2-mj+j)\phi}{2}\right]\,S_{j,m+1}
+2\,\sum_{j=0}^{m-1}\cos\left(\frac{j\phi}{2}\right)S_{m-j,m-j}\,S_{j,j+1} 
\nonumber \\ 
& &\mbox{}+2\,\sum_{j=1}^{m-1}\sum_{k=0}^{j-1}S_{m-j,m-k}\left\{
\cos\left[\frac{(mj-mk-j)\phi}{2}\right]\,S_{j,k+1}
+\cos\left[\frac{(mj-mk+k)\phi}{2}\right]\,S_{k,j+1}\right\}. 
\end{eqnarray}
The explicit expressions for the first few $P_{m,m}$ are 
(with $\theta=\phi/2$):
\begin{eqnarray*}
P_{1,1}&=&14\cos\theta+2\cos3\theta, \\
P_{2,2}&=&26+32\cos2\theta+26\cos4\theta+4\cos6\theta+2\cos8\theta, \\
P_{3,3}&=&130\cos\theta+124\cos3\theta+88\cos5\theta+52\cos7\theta
+40\cos9\theta+8\cos11\theta+4\cos13\theta+2\cos15\theta, \\
P_{4,4}&=&224+410\cos2\theta+396\cos4\theta+308\cos6\theta+282\cos8\theta
+188\cos10\theta+130\cos12\theta \\  
& &\mbox{}+76\cos14\theta+58\cos16\theta+14\cos18\theta+8\cos20\theta
+4\cos22\theta+2\cos24\theta, \\
P_{5,5}&=&1446\cos\theta+1386\cos3\theta+1308\cos5\theta+1176\cos7\theta
+1032\cos9\theta+842\cos11\theta+690\cos13\theta \\
& &\mbox{}+542\cos15\theta+398\cos17\theta+264\cos19\theta
+180\cos21\theta+108\cos23\theta+80\cos25\theta \\
& &\mbox{}+24\cos27\theta
+14\cos29\theta+8\cos31\theta+4\cos33\theta+2\cos35\theta, \\
P_{6,6}&=&2518+4868\cos2\theta+4808\cos4\theta+4514\cos6\theta+4238\cos8\theta
+3788\cos10\theta+3466\cos12\theta \\  
& &\mbox{}+2938\cos14\theta+2554\cos16\theta+2074\cos18\theta
+1702\cos20\theta+1298\cos22\theta+1056\cos24\theta \\
& &\mbox{}+736\cos26\theta+536\cos28\theta
+356\cos30\theta+244\cos32\theta+148\cos34\theta+110\cos36\theta \\
& &\mbox{}+38\cos38\theta+24\cos40\theta+14\cos42\theta
+8\cos44\theta+4\cos46\theta+2\cos48\theta.
\end{eqnarray*}
These $P_{m,m}$'s are plotted in Fig.~4. Note that $P_{m,m}$ depends only on 
the even (odd) harmonics of
$\theta$ and has a period $2\pi$ ($4\pi$) for even (odd) $m$. 
The expressions for $P_{m,m}$ are 
obviously more complicated than the corresponding $I_{2m}$. 
However, by comparing Figs.~3 and 4, we find that the general features in 
the interference behaviors are surprisingly similar. We thus infer that 
the relevant physical quantities are not significantly changed 
by the addition of interference between the dominant winding paths.

\section{Effects of disorder}
\subsection{Average of the tunneling probability}  
To incorporate the effects of random impurities, we now replace the 
on-site energy part in Eq.~(1) (first term in $H$)
by $\sum_{i}\epsilon_{i}c_{i}^{\dag}c_{i}$, where the $\epsilon_{i}$'s 
are now independent random variables. The Hamiltonian now takes the form 
$$H=\sum_{i} \epsilon_{i}\,c_{i}^{\dag}c_{i} + V \sum_{\langle ij \rangle}
c_{i}^{\dag}c_{j} \exp(i A_{ij})\,.$$
We have studied two commonly used 
models: (i) $\epsilon_{i}$ can take two values: $+W$ and $-W$ with equal 
probability; and (ii) $\epsilon_{i}$ is randomly chosen from a uniform 
distribution of width $W$ and zero mean. We found that both models 
yield  the same results for the MC.

We now start with the  general case of the first model, namely,  
$\epsilon_{i}$ can have two values: $+W$ with
probability $\mu$ and $-W$ with probability $\nu$, where $\mu+\nu=1$. 
Due to disorder, the transition amplitude becomes 
$$T_{if}=W\left(\frac{V}{W}\right)^rJ_{m,n}\,,$$ with 
\begin{equation}
J_{m,n}=\sum_{\Gamma}\left[\prod_{j\in \Gamma}\left(
-\frac{W}{\epsilon_j}\right)\right]e^{i\Phi_{\Gamma}},
\end{equation}
where $\Gamma$ runs over all directed paths of $r$ steps connecting sites 
$(0,0)$ and $(m,n)$, and $j$ over sites on each path. 
For all directed paths ending at $(m,n)$, electrons traverse $r=m+n$
sites (the initial site $(0,0)$ is excluded). Each site visited now
contributes an additional multiplicative factor of either $+1$ or $-1$
to the phase factor. Therefore, for a given path $\Gamma$, the
probability for obtaining 
$\ \pm e^{i\Phi_{\Gamma}} \ $ is 
$$P_{\pm}=\frac{(\mu+\nu)^{r}\pm(\mu-\nu)^{r}}{2}.$$ 
It is then clear that
\begin{equation}
\langle\,J_{m,n}(\phi)\,\rangle=(P_{+}-P_{-})\,S_{m,n}(\phi)
=(\mu-\nu)^{r}\,S_{m,n}(\phi),
\end{equation}
where $\langle\cdots\rangle$ denotes averaging over all impurities. 

By exploiting Eqs.~(7) and (8), we derive the following general 
expressions valid for {\em any} $\mu$ and $\nu$,
\begin{equation}
\langle\,J^{2}_{m,n}(\phi)\,\rangle \,=\,(1-{\cal P})\,S_{m,n}(2\phi)
\,+\,{\cal P}\,S^{2}_{m,n}(\phi),
\end{equation}
where 
\begin{eqnarray}
{\cal P}&=&P_{+}^{N}+P_{-}^{N}
+\sum_{k=1}^{N-1}\,P_{+}^{N-k}\,P_{-}^{k}\,(C_{k}^{N}-4\,C^{N-2}_{k-1}) 
\nonumber \\
&=&1-4\,P_{+}\,P_{-}=(\mu-\nu)^{2r}.
\end{eqnarray}
Also, the disorder average of the tunneling probability 
(i.e., the transmission rate) $|J|^2=JJ^*$ is 
\begin{equation}
\langle\,|J_{m,n}(\phi)|^{2}\,\rangle \,=\,(1-{\cal P})\,N\,
+\,{\cal P}\,S^{2}_{m,n}(\phi).
\end{equation}

The physical origin of Eqs.~(29) and (31) becomes clearer by 
rewriting them as
\begin{equation}
\langle\,J^{2}_{m,n}(\phi)\,\rangle = S_{m,n}(2\phi)
+{\cal P}\,S_{m,n}(\phi)\left[\,S_{m,n}(\phi)-C_{m,n}(\phi)\right]
\end{equation}
and
\begin{equation}
\langle\,|J_{m,n}(\phi)|^{2}\,\rangle = N
+{\cal P}\,[\,S^{2}_{m,n}(\phi)-N],
\end{equation}
where 
$$
C_{m,n}(\phi)=\frac{\prod_{k=1}^{m+n}\cos\frac{k}{2}\phi}
{\left(\,\prod_{k=1}^{m}\cos\frac{k}{2}\phi\right)
\left(\,\prod_{k=1}^{n}\cos\frac{k}{2}\phi\right)}\,,
$$
and we have used 
$$S_{m,n}(2\phi)=S_{m,n}(\phi)C_{m,n}(\phi).$$ 
The first terms in Eqs.~(32) and (33) account for contributions 
from pairs of identical paths: 
$$\sum_{\Gamma}\left(\pm e^{i\Phi_{\Gamma}}\right)
\left(\pm e^{i\Phi_{\Gamma}}\right)
=\sum_{\Gamma} e^{2i\Phi_{\Gamma}}=S_{m,n}(2\phi)$$ 
in $\langle \,J^{2}_{m,n}\,\rangle$, and 
$$\sum_{\Gamma}\left(\pm e^{i\Phi_{\Gamma}}\right)
\left(\pm e^{-i\Phi_{\Gamma}}\right)=\sum_{\Gamma}1=N$$ 
in $\langle \,|J_{m,n}|^{2}\,\rangle$. 
The second terms in Eqs.~(32) and (33) account for contributions 
from pairs of distinct paths.
Note that $S_{m,n}(0)=N$  and $C_{m,n}(0)=1$, when $\phi=0$. We then have 
in the absence of magnetic flux  
\begin{equation}
\langle\,J^{2}_{m,n}(0)\,\rangle =
\langle\,|J_{m,n}(0)|^{2}\,\rangle =
N+{\cal P}\,N\,(N-1).
\end{equation}
Furthermore, in the special case $\mu=\nu=1/2$, 
since ${\cal P}=0$ we then obtain
\begin{eqnarray*}
\langle\,J_{m,n}(\phi)\,\rangle &=&0, \\ 
\langle\,J^{2}_{m,n}(\phi)\,\rangle &=&S_{m,n}(2\phi), \\ 
\langle\,|J_{m,n}(\phi)|^{2}\,\rangle &=&N.
\end{eqnarray*} 

\subsection{Higher-order moments and general expressions for the first 
few leading terms}
For $\mu=\nu=1/2$ (the most studied case so far), 
we can obtain analytical expressions for the moments 
$ \langle \, J^{2p}_{m,n}(\phi) \, \rangle $ and
$ \langle \, |J_{m,n}(\phi)|^{2p} \, \rangle $ 
for any value of $p$. Only a few of these will be presented here. 
From now on, $J(\phi)$ stands for $J_{m,n}(\phi)$ and 
$S(\phi)$ stands for $S_{m,n}(\phi)$. 
The derivation of these moments is given in appendix A.

\setcounter{equation}{0}
\renewcommand{\theequation}{35.\arabic{equation}}
\begin{eqnarray}
\langle \, J^{4}(\phi) \, \rangle& =&
3S^{2}(2\phi)\,-\,2\,S(4\phi), \\
\langle \, J^{6}(\phi) \, \rangle &=&
15\,S^{3}(2\phi)-30\,S(2\phi)\,S(4\phi)+16\,S(6\phi), \\
\langle \, J^{8}(\phi) \, \rangle &=&
105\,S^{4}(2\phi)-\,420\,S^{2}(2\phi)\,S(4\phi)+448\,S(2\phi)\,S(6\phi)
+140\,S^{2}(4\phi)-272\,S(8\phi), \\
\langle \, J^{10}(\phi) \, \rangle &=&
945\,S^{5}(2\phi)-6300\,S^{3}(2\phi)\,S(4\phi)+10080\,S^{2}(2\phi)\,S(6\phi)
+6300\,S(2\phi)\,S^{2}(4\phi) \nonumber \\
& &\mbox{}-12240\,S(2\phi)\,S(8\phi)-6720\,S(4\phi)\,S(6\phi)+7936S(10\phi), \\
\langle \, J^{12}(\phi) \, \rangle &=&
10395\,S^{6}(2\phi)-103950\,S^{4}(2\phi)\,S(4\phi)
+221760\,S^{3}(2\phi)\,S(6\phi)+207900\,S^{2}(2\phi)\,S^{2}(4\phi) \nonumber \\
& &\mbox{}-403920\,S^{2}(2\phi)\,S(8\phi)
-443520\,S(2\phi)\,S(4\phi)\,S(6\phi)-46200\,S^{3}(4\phi)  \nonumber \\
& &\mbox{}+523776\,S(2\phi)\,S(10\phi)+269280\,S(4\phi)\,S(8\phi)
+118272\,S^{2}(6\phi)-353792\,S(12\phi), \\ 
\langle \, J^{14}(\phi) \, \rangle &=&
135135\,S^{7}(2\phi)-1891890\,S^{5}(2\phi)\,S(4\phi)
+5045040\,S^{4}(2\phi)\,S(6\phi)+6306300\,S^{3}(2\phi)\,S^{2}(4\phi) 
\nonumber \\
& &\mbox{}-12252240\,S^{3}(2\phi)\,S(8\phi)
-20180160\,S^{2}(2\phi)\,S(4\phi)\,S(6\phi)
-4204200\,S(2\phi)\,S^{3}(4\phi)  \nonumber \\
& &\mbox{}+23831808\,S^{2}(2\phi)\,S(10\phi)
+24504480\,S(2\phi)\,S(4\phi)\,S(8\phi)
+10762752\,S(2\phi)\,S^{2}(6\phi)  \nonumber \\
& &\mbox{}+6726720\,S^{2}(4\phi)\,S(6\phi)
-32195072\,S(2\phi)\,S(12\phi)-15887872\,S(4\phi)\,S(10\phi) \nonumber \\
& &\mbox{}-13069056\,S(6\phi)\,S(8\phi)+22368256\,S(14\phi),
\end{eqnarray}
and
\setcounter{equation}{0}
\renewcommand{\theequation}{36.\arabic{equation}}
\begin{eqnarray}
\langle \, |J(\phi)|^4 \, \rangle &=&2\,N\,(N-1)+S^{2}(2\phi), \\
\langle \,|J(\phi)|^6\,\rangle &=&2\,N\,(3N^2-9N+8)
+3\,(3N-4)\,S^{2}(2\phi), \\ 
\langle \, |J(\phi)|^8 \, \rangle &=&
8\,N\,(3N^3-18N^2+41N-34)+8\,(9N^2-33N+32)\,S^{2}(2\phi)
\nonumber \\ 
& &\mbox{}+9\,S^{4}(2\phi)-12\,S^{2}(2\phi)\,S(4\phi)+4\,S^{2}(4\phi), \\
\langle \, |J(\phi)|^{10} \, \rangle &=&
8\,N\,(15N^4-150N^3+625N^2-1250N+992)+
40\,(15N^3-105N^2+260N-216)\,S^{2}(2\phi) \nonumber \\ 
& &\mbox{}+75\,(3N-8)\,S^{4}(2\phi)-20\,(15N-44)\,S^{2}(2\phi)\,S(4\phi)
+20\,(5N-16)\,S^{2}(4\phi), \\
\langle \, |J(\phi)|^{12} \, \rangle &=&
16\,N\,(45N^5-675N^4+4425N^3-15525N^2+28706N-2212) \nonumber \\
& &\mbox{}+120\,(45N^4-510N^3+2295N^2-4702N+3552)\,S^{2}(2\phi)
+30\,(135N^2-854N+1440)\,S^{4}(2\phi) \nonumber \\ 
& &\mbox{}+225\,S^{6}(2\phi)-120\,(45N^2-309N+556)\,S^{2}(2\phi)\,S(4\phi)
-900\,S^{4}(2\phi)\,S(4\phi) \nonumber \\
& &\mbox{}+24\,(75N^2-555N+1064)\,S^{2}(4\phi)
+900\,S^{2}(2\phi)\,S^{2}(4\phi)+480\,S^{3}(2\phi)\,S(6\phi) \nonumber \\
& &\mbox{}-960\,S(2\phi)\,S(4\phi)\,S(6\phi)+256\,S^{2}(6\phi), \\
\langle \, |J(\phi)|^{14} \, \rangle &=&
16\,N\,(315N^6-6615N^5+62475N^4-334425N^3+1057322N^2-1854160N+1398016) 
\nonumber \\
& &\mbox{}+56\,(945N^5-1575N^4+110775N^3-401730N^2+732536N-518464)
\,S^{2}(2\phi)
 \nonumber \\
& &\mbox{}+1470\,(45N^3-495N^2+1920N-2576)\,S^{4}(2\phi)
+11025\,(N-4)\,S^{6}(2\phi) \nonumber \\ 
& &\mbox{}-840\,(105N^3-1239N^2+5096N-7192)\,S^{2}(2\phi)\,S(4\phi)
-2940\,(15N-64)\,S^{4}(2\phi)\,S(4\phi) \nonumber \\ 
& &\mbox{}+2940\,(15N-68)\,S^{2}(2\phi)\,S^{2}(4\phi)
+3360\,(7N-31)\,S^{3}(2\phi)\,S(6\phi)  \nonumber \\
& &\mbox{}+56\,(525N^3-6615N^2+28784N-42688)\,S^{2}(4\phi)
-448\,(105N-493)\,S(2\phi)\,S(4\phi)\,S(6\phi) \nonumber \\
& &\mbox{}+1792\,(7N-34)\,S^{2}(6\phi).
\end{eqnarray}
These moments satisfy the consistency check 
$\langle \, J^{2p}(0)\, \rangle = \langle \, |J(0)|^{2p} \, \rangle$ 
and odd moments vanish by symmetry.

The moments provide an analytical view of the structure of the 
QI in the tunneling process. Since $|J|^2=JJ^*$,
each $|J|^2$ represents $N$ {\em forward} paths to $(m,n)$, 
each one with its corresponding {\em reversed} path back to the origin.  
\ Also, 
$\, \langle \, |J(\phi)|^{2p} \, \rangle , \, $
averages over the contributions of $N^p$ such pairs of paths. 
In general, $\langle\, |J(\phi)|^{2p} \,\rangle$ consist of terms involving 
$N^{k} (k=1, \cdots, p).$ The above explicit expressions for the moments 
will allow us to deduce general formulae for the first few leading 
(i.e., dominant) terms in the moments. 

We first focus on the leading terms 
($\propto N^{p}$), since they provide the most significant contribution to 
the moments when $N$ is large. Recall that $S(0)=N$, therefore we need to 
consider all terms involving 
$S^{2k}(2\phi)\,N^{p-2k}$ in $\langle\, |J(\phi)|^{2p} \,\rangle$. 
We derive (see appendix~B for more details)
\setcounter{equation}{36}
\renewcommand{\theequation}{\arabic{equation}}
\begin{eqnarray}
\langle\, |J(0)|^{2p} \,\rangle &=&(2p-1)!!\,N^{p}, \\
\langle\, |J(\phi)|^{2p} \,\rangle &=& p!\,N^{p}\,\left\{\sum_{k=0}^{\infty}
\frac{(2k)!\,C^{p}_{2k}}{(2^k\,k!)^{2}}\,
\left[\frac{S(2\phi)}{N}\right]^{2k}\right\}. 
\end{eqnarray}
Furthermore, by considering all the second leading terms ($\propto N^{p-1}$), 
we obtain
\begin{eqnarray}
\langle\, |J(0)|^{2p} \,\rangle &=&-\frac{1}{3}\,p\,(p-1)\,[(2p-1)!!]
\,N^{p-1}, \\
\langle\, |J(\phi)|^{2p} \,\rangle &=&-\frac{p!\,N^{p-1}}{6}\,
\left\{\sum_{k=0}^{\infty}
\frac{(2k+1)!}{(2^k\,k!)^{2}}\left[ (3p+2k-3)\,C^{p}_{2k+1}
+2k\,C^{p}_{2k+2}\frac{S(4\phi)}{N}\right]\left[
\frac{S(2\phi)}{N}\right]^{2k}\right\}. 
\end{eqnarray}
Also, when $S(2p\phi)=0$ with $p \geq 1$ (e.g., at 
$\phi=\pi/m$, $S_{m,m}(2p\phi)=0$), 
the third leading terms ($\propto N^{p-2}$) in the moments are
\begin{eqnarray}
\langle\, |J(0)|^{2p} \,\rangle &=&\frac{1}{90}\,p\,(p-1)\,(p-2)\,(5p+1)
\,[(2p-1)!!]\,N^{p-2}\,, \\
\langle\, |J(\phi)|^{2p} \,\rangle &=&\frac{1}{72}\,p\,(p-1)\,(p-2)
\,(9p+5)\,(p!)\,N^{p-2}\,.
\end{eqnarray}
The above general expressions for the moments are of value 
since they  enable us to {\em analytically} obtain the 
dominant contributions to the quantity we are interested in: 
the magnetoconductance.

\subsection{Analytical results for the magnetoconductance}
We now use the replica method: 
\begin{equation}
\langle\, \ln|J(\phi)|^{2} \,\rangle=\lim_{p\rightarrow 0}
\frac{\langle\, |J(\phi)|^{2p} \,\rangle -1}{p}
\end{equation}
to compute the log-averaged MC with respect to the zero-field log-averaged MC 
(denoted by $L_{MC}$), defined as 
\begin{eqnarray}
L_{MC}& \equiv& \langle\, \ln|J(\phi)|^{2} \,\rangle 
-\langle\, \ln|J(0)|^{2} \,\rangle  \nonumber \\
&=& \lim_{p\rightarrow 0} \frac{\langle\, |J(\phi)|^{2p} \,\rangle - 
\langle\, |J(0)|^{2p} \,\rangle}{p}\,.
\end{eqnarray}
Taking into account only the first leading terms in the moments, shown in 
Eqs.~(37) and (38), we derive the $L_{MC}$ as 
\begin{equation}
L_{MC}=\ln 2-\sum_{k=1}^{\infty}\frac{(2k-1)!}{(2^k\,k!)^2}
\left[\frac{S(2\phi)}{N}\right]^{2k},
\end{equation} 
where we have
\begin{equation}
\sum_{k=1}^{\infty}\frac{(2k-1)!}{(2^k\,k!)^2}=\ln2\,.
\end{equation} 
Exploiting the following identity\cite{table} for $0<x\leq 1$ 
$$
\cosh^{-1}\frac{1}{x}=\ln\frac{2}{x}
-\sum_{k=1}^{\infty}\frac{(2k-1)!}{(2^k\,k!)^2}x^{2k}\,,
$$
which reduces to Eq.~(46) for $x=1$, we thus obtain a very concise exact 
expression for the $L_{MC}$ as
\begin{eqnarray}
L_{MC}&=&\left\{ 
\begin{array}{ll}
\cosh^{-1}\frac{N}{\left|S(2\phi)\right|}-\ln\frac{N}{\left|S(2\phi)\right|} &
\mbox{when $S(2\phi)\neq 0$} \\
\ln 2 & \mbox{when $S(2\phi)=0$} 
\end{array}
\right. \\
&=&\ln\left[1+\sqrt{1-\left(\frac{S(2\phi)}{N}\right)^2}\,\right].
\end{eqnarray}

The typical MC of a sample 
$$G(\phi)=\exp(\langle \ln |J(\phi)|^2\rangle)$$  
is then given by, normalized by the zero-field MC $G(0)$,
\begin{equation}
\frac{G(\phi)}{G(0)}=\exp(L_{MC})=1
+\sqrt{1-\left[\frac{S(2\phi)}{N}\right]^2}\,.
\end{equation}
Eq.~(49) is one of our main results.  It provides a concise 
closed-form expression 
for the MC, as an {\em explicit} function of the magnetic flux.
From Eq.~(49) it becomes evident that 
a magnetic field leads to an increase in the {\em positive} MC: 
$G(\phi)/G(0)$ increases from $1$ to a saturated value $2$ 
(since $S(2\phi)$ decreases from $N$ to $0$) when the flux 
is turned on and increased. $G(\phi)=2G(0)$ at the field $\phi$ that satisfies 
$S(2\phi)=0$. Furthermore, 
it is clear that the MC varies {\em periodically} with the 
magnetic field and the 
periodicity in the flux is equal to the superconducting flux quantum 
$hc/2e$.

It is important to point out that Eqs.~(48) and (49) are valid in any 
dimension as long as we use the corresponding D-dimensional sum $S^{(r)}$.

It is illuminating to draw attention to the 
close relationship between the behaviors of 
$I_{2m}(2\phi)=S_{m,m}(2\phi)$ and the corresponding $G(\phi)$. When 
$\phi=0$, $(I_{2m}(0)/N)^2=1$, which is the {\em largest} value 
of $(I_{2m}(2\phi)/N)^2$ as a function of $\phi$, and the MC is 
equal to the {\em smallest} value $G(0)$. When the magnetic field is 
increased from zero, $(I_{2m}(2\phi)/N)^2$ quickly approaches 
(more rapidly as $m$ becomes larger) its {\em smallest} 
value, which is zero, at $\phi/2\pi=1/4m$. 
At the same time, the MC 
rapidly increases to the {\em largest} value $2G(0)$.
 
The physical implication of this is clear: fully constructive 
interference in the case without disorder leads to  the 
smallest hopping conduction in the presence of disorder. While fully 
destructive interference in the case without disorder yields the 
largest hopping conduction in the presence of disorder. 
Moreover, when $m$ (the system size is $m\times m$) is large, 
$G(\phi)/G(0)$ remains in the close vicinity of $2$ for 
$\phi/2\pi>1/4m$ in spite of the strong very-small-magnitude 
fluctuations of $I_{2m}(2\phi)/N$ around zero. 

The saturated value of the 
magnetic field $B_{{\rm sa}}$ (i.e., the first field that 
makes $G(\phi)=2G(0)$) is 
inversely proportional to twice the hopping length: the larger the system is, 
the smaller $B_{{\rm sa}}$ will be. In other words, as soon as the 
system, with hopping distance $r=2m$, is penetrated by a total flux 
of $(1/2r)\times(r/2)^2=r/8$ (in units of the 
flux quantum $hc/e$), 
the MC reaches the saturation value $2G(0)$.  

Defining the relative MC, $\Delta G(\phi)$, as
$$\Delta G(\phi)\equiv\frac{G(\phi)-G(0)}{G(0)}\,,$$
and utilizing Eq.~(49), we show $\Delta G(\phi)$ versus $\phi$ for several 
different hopping lengths in Fig.~5. 
The behavior of $\Delta G(\phi)$ described above 
can be clearly observed in these figures.

Now let us examine the behavior of the MC in the low-flux limit. 
From Eqs.~(14) and (15), 
it follows then that, for very small fields, in 2D
\begin{equation}
\Delta G(\phi) \simeq \frac{\sqrt{3}}{6}\,r^{3/2}\,\phi,
\end{equation} 
and in ladder-type quasi-1D structures
\begin{equation}
\Delta G(\phi) \simeq \frac{\sqrt{3}}{3}\,r\,\phi.
\end{equation} 
In Fig.~6, we plot $\Delta G(\phi)$ computed directly from Eq.~(49), 
for various small values of $\phi$, versus $r^{3/2}\phi$, 
with $r=2, 4,\ldots, 1000$, in (a) 
and versus $r\phi$, with $r=2, 3,\ldots, 500$, in (b), 
respectively for 2D and quasi-1D systems. 
It is seen that, both in (a) and (b), all the data nicely collapses into 
a straight line, which verifies the scaling of the low-flux MC in 
Eqs.~(50) and (51).

If we consider the second leading terms in the moments, 
namely Eqs.~(39) and (40), the second-order contribution to the 
$L_{MC}$ is
\begin{eqnarray}
L_{MC}&=&\left\{
\begin{array}{ll}
0 & \mbox{when $S(2\phi)/N=\pm 1$} \\
\frac{1}{6N}\,\left\{1-\sum_{k=1}^{\infty}
\left[\frac{(2k+1)!}{(2^k\,k!)^{2}}\left(\frac{2k-3}{2k+1}-\frac{k}{k+1}
\frac{S(4\phi)}{N}\right)\left(\frac{S(2\phi)}{N}\right)^{2k}
\right]\right\} &
\mbox{when $S(2\phi)/N\neq \pm 1$}
\end{array}
\right. \nonumber \\
&=&\left\{
\begin{array}{ll}
0 & \mbox{when $\Lambda=0$} \\
\frac{1}{6N}\,\left\{1-\frac{1-\Lambda}{\Lambda^3(1+\Lambda)}
\left[\left(1-\frac{S(4\phi)}{N}\right)
(1+2\Lambda)-\Lambda^2(2+3\Lambda)\right]\right\} &
\mbox{when $\Lambda\neq 0$}
\end{array}
\right.,
\end{eqnarray}
where
$$\Lambda=\sqrt{1-\left(\frac{S(2\phi)}{N}\right)^2}\,,$$
and we have used
\begin{eqnarray*}
\sum_{k=1}^{\infty}\frac{(2k)!}{(2^kk!)^2}\,x^{2k} 
& =& \frac{1-\sqrt{1-x^2}}{\sqrt{1-x^2}}\,, \\
\sum_{k=1}^{\infty}2k\frac{(2k)!}{(2^kk!)^2}\,x^{2k} 
& =& \frac{x^2}{(1-x^2)^{3/2}}\,, \\
\sum_{k=1}^{\infty}\frac{(2k)!}{(k+1)(2^kk!)^2}\,x^{2k} 
& =& \frac{x^2}{(\,1-\sqrt{1-x^2}\,)^2}\,. 
\end{eqnarray*}
The principal features in the behavior of the MC are not significantly 
modified by the addition of the contribution from $L_{MC}$ in Eq.~(52): 
while the magnitude of $\Delta G(\phi)$ is slightly increased for 
$\phi\neq 0$, the period of the MC remains unchanged.

For a 2D system and in the low-flux limit, we derive from Eq.~(52) 
for diagonal sites $(r/2,r/2)$:  
$$\Delta G(\phi)=(\sqrt{3}/24N)\,r^{3/2}\,\phi.$$ 
Comparing this result with 
Eq.~(50), we see that the dependence of the small-field MC on the hopping 
length and the field is the same except for different prefactors. 
Summing up both contributions, we have for small $\phi$ 
\begin{equation}
\Delta G(\phi)=\frac{\sqrt{3}}{6}\left(1+\frac{1}{4N}\right)r^{3/2}\,\phi\,.
\end{equation}

In addition, when $S(2p\phi)=0$, we have from Eqs.~(37-42)
\begin{equation}
L_{MC}=\ln 2+\frac{1}{6\,N}+\frac{7}{60\,N^{2}}
+O\left(\frac{1}{N^{3}}\right).
\end{equation}
This indicates that the magnitude of the positive MC is gradually 
increased (e.g., the saturation value of $\Delta G$ 
is raised above $1$) when contributions from higher-order 
terms (i.e., terms $\propto 1/N^k$ with $k\geq 1$) are included, 
though they are negligibly small.

\subsection{Discussion}
Our results for the MC are in good agreement with experimental 
measurements.\cite{exp1,exp2} For instance, a positive MC is observed in 
the VRH regime of both macroscopically large In$_2$O$_{3-x}$ 
samples\cite{exp1} and compensated n-type CdSe.\cite{exp2} 
Moreover, saturation in 
the MC as the field is increased is also reported in Ref.~20. 

The results for $\Delta G(\phi)$ presented in this work are consistent with 
theoretical studies based on an independent-directed-path 
formalism\cite{7} 
and a random matrix theory of the transition strengths.\cite{rmt} 
The advantages of our results include: (i) they provide explicit 
expressions for the first two dominant contribution to the MC, 
as a function of the magnetic field; 
(ii) they provide straightforward determination of the period of 
the oscillation of the MC; (iii) they provide explicit scaling behaviors 
(i.e., the dependence on the hopping 
length and the orientation and strength of the field) 
of the low-flux MC in quasi 1-D, 2D and 
3D systems; and (iv) they allow us to make quantitative 
comparison with experimental data. Finally, it is important to 
emphasize that our analytic results 
[Eqs.~(48), (49) and (52)] for the MC are 
equally applicable to {\em any dimension}, since the essential 
ingredient in our expressions is the QI quantity $S^{(r)}$, which 
takes into account the dimensionality.

In appendix B, 
we outline the computational scheme using the second model 
of disorder, i.e., $\epsilon_i$ is randomly chosen from the interval 
$[-W/2,W/2]$. The moments obtained in this case 
are the same as those presented in Eqs.~(37) and (38). 
Therefore, the result for the MC remains unchanged.

\section{Quantum Interference and the small-field magnetoconductance 
on a three-dimensional cubic lattice}
\subsection{Sums over forward-scattering paths} 
Let ${\cal S}_{m,n,l}$ 
($=S^{(r)}$ in 3D) be the sum over all phase factors associated 
with directed paths of $m+n+l \, (=r)$ steps 
along which an electron may hop from (0,0,0) to the site $(m,n,l)$. 
Again we assume $m$, $n$, and $l \geq 0$. In other words, electrons 
can now also hop in the positive $z$ direction. The vector potential of a 
general magnetic field $(B_{x},B_{y},B_{z})$ can be written as 
$${\bf A}=\frac{1}{2}(zB_y-yB_z, \, xB_z-zB_x, \, yB_x-xB_y ).$$  Also, 
$a/2\pi$, $b/2\pi$ and $c/2\pi$ represent the three fluxes through the 
respective elementary plaquettes on the $yz$-, $zx$- and $xy$-planes. 
To compute ${\cal S}_{m,n,l}$, we start from the following 
recursion relation
\begin{equation}
{\cal S}_{m,n,l}=\sum_{p=0}^{m} 
\sum_{q=0}^{n}A_{p,q,l\rightarrow m,n,l}\,\exp\left(i
\frac{qa-pb}{2}\right)\,{\cal S}_{p,q,l-1},
\end{equation}
where $A_{p,q,l\rightarrow m,n,l}$ is the sum over all directed 
paths starting from $(p,q,l)$ and ending at $(m,n,l)$.
The physical meaning of Eq.~(55) 
is as follows. The site $(m,n,l)$ is 
reached by taking one step from $(p,q,l-1)$ to $(p,q,l)$, acquiring 
the phase $i(qa-pb)/2$, then traversing from $(p,q,l)$ to $(m,n,l)$ 
on the $z=l$ plane. After some calculation, we find that
\begin{equation}
A_{p,q,l\rightarrow m,n,l}=\exp\left\{i\left
[\frac{(m-p)(lb-qc)+(n-q)(pc-la)}{2}
\right]\right\}\,S_{m-p,n-q}(c),
\end{equation}
where $ S_{m-p,n-q}(c) $ is defined as shown in Eq.~(10). By 
applying Eq.~(56) 
$l$ times, we obtain a general formula 
of ${\cal S}_{m,n,l}$ for $m,n,l \geq 1$ 
in terms of the fluxes $a$, $b$ and $c$ as

\begin{equation}
{\cal S}_{m,n,l}(a,b,c)=\exp\left[-i\left(\frac{nla+lmb+mnc}{2}\right)
\right]\,{\cal L}_{m,n,l}(a,b,c),
\end{equation}
where
\begin{equation}
{\cal L}_{m,n,l}(a,b,c)=\left\{\prod_{j=1}^{l}
\left[ \sum_{p_j=0}^{p_{j+1}}\sum_{q_j=0}^{q_{j+1}}\exp\left\{i[q_ja+(m-p_j)
b+p_j(q_{j+1}-q_j)c]\right\}\,L_{p_{j+1}-p_j,q_{j+1}-q_j}(c)\right]\right\}
\,L_{p_1,q_1}(c), 
\end{equation}
with $p_{l+1}\equiv m$, $q_{l+1}\equiv n$, and the $L_{p,q}(c)$'s 
are defined as in Eq.~(12). 

It is clear that ${\cal S}_{m,n,0}=S_{m,n}(c)$, 
${\cal S}_{m,0,l}=S_{m,l}(b)$, and ${\cal S}_{0,n,l}=S_{n,l}(a)$. 
Also, the following symmetries hold:
\begin{equation}
{\cal S}_{m,n,l}(a,b,c)={\cal S}_{n,m,l}(b,a,c)={\cal S}_{m,l,n}(a,c,b)
={\cal S}_{l,m,n}(b,c,a)={\cal S}_{l,n,m}(c,b,a)
={\cal S}_{n,l,m}(c,a,b).
\end{equation} 
When there is no magnetic flux, 
$${\cal S}_{m,n,l}(0,0,0)=\frac{(m+n+l)!}{m!\,n!\,l!} \equiv {\cal N}$$ 
gives the total number of $(m+n+l)$-step paths connecting $(0,0,0)$ and 
$(m,n,l)$.

We have obtained explicit expressions for many ${\cal S}_{m,n,l}$, and 
here we explicitly present only the first few ${\cal S}$, since ${\cal S}$ 
have long expressions for larger $m$, $n$, and $l$.
\begin{eqnarray*}
{\cal S}_{1,1,1}&=&2\left[\cos\frac{a+b-c}{2}+\cos\frac{b+c-a}{2}
+\cos\frac{c+a-b}{2}\right], \\
{\cal S}_{2,1,1}&=&2\left[\cos\frac{a}{2}+\cos\left(\frac{a}{2}-b\right)
+\cos\left(\frac{a}{2}-c\right)+\cos\left(\frac{a}{2}\pm b\mp c\right)
+\cos\left(\frac{a}{2}-b-c\right)\right], \\
{\cal S}_{2,2,1}&=&2+2\sum\!{ ^{(1)}}\cos\alpha+4\cos(a-b)+2\cos(a-c)
+2\cos(b-c)+2\cos(a-2c)+2\cos(b-2c) \\
& &\mbox{}+2\cos(a-b\pm c)+2\cos(a+b-2c)+2\cos(a-b\pm 2c), \\
{\cal S}_{2,2,2}&=&6+\sum\!{ ^{(1)}}\left[4\cos\alpha+2\cos2\alpha\right]
+\sum\!{ ^{(2)}}\left[4\cos(\alpha-\beta)+4\cos2(\alpha-\beta)
+2\cos(\alpha-2\beta)+2\cos(2\alpha-\beta)\right] \\
& &\mbox{}+2\sum\!{ ^{(3)}}\left[\cos(\alpha+\beta-\gamma)
+\cos2(\alpha+\beta-\gamma)+\cos(\alpha+\beta-2\gamma)
+\cos(\alpha\pm 2\beta\mp 2\gamma)\right].
\end{eqnarray*}
Here $\sum\!{ ^{(i)}}$ denote sums over $\alpha=a, b, c$;
$(\alpha \, \beta)=(a\,b), (b\,c), (c\,a)$; and 
$(\alpha \, \beta \, \gamma)=(a\,b\,c), (b\,c\,a), (c\,a\,b)$;
for $i=1$, $2$, and $3$ respectively; and, for instance, 
the term\  $\cos(a/2 \pm b\mp c)$\  means\  $\cos(a/2+b-c)+\cos(a/2-b+c)$.

\subsection{Low-flux limit}
In the very-low-flux limit, and calculated exactly to second-order 
in the flux, we obtain the logarithm of ${\cal S}_{m,n,l}$, the 3D analog 
of the harmonic shrinkage of the wave function, as
\begin{equation}
\ln{\cal S}_{m,n,l}=\ln{\cal N}-\frac{1}{24}\left[nla^2+lmb^2+mnc^2
+m(lb-nc)^2+n(mc-la)^2+l(na-mb)^2\right]. \nonumber
\end{equation}
This generalizes the 2D harmonic-shrinkage of the wave function 
obtained in Eq.~(13).
When $m=n=l=r/3$, we have
\begin{equation}
\ln{\cal S}_{m,m,m}=\ln \frac{r!}{[(r/3)!]^3}-\frac{1}{216}
\left\{r^{2}(a^2+b^2+c^2)
+\frac{r^{3}}{3} \left[(b-c)^2+(c-a)^2+(a-b)^2\right]\right\}. \nonumber
\end{equation}
These results generalize to 3D the 2D results obtained in Sec.~II~B.

\subsection{Interference patterns on diagonal sites}
In order to see how the interference patterns vary according to 
the orientation of the applied field, 
we focus on ${\cal S}_{m,m,m}$ (i.e., ${\cal S}$ on the body diagonals). 
We now examine two special cases: 
${\bf B}_{\parallel}\equiv {\bf B} \parallel (1,1,1)
=(\phi,\phi,\phi)/2\pi$ and 
${\bf B}_{\perp}\equiv {\bf B} \perp (1,1,1)=
(\phi/2,\phi/2,-\phi)/2\pi$, 
namely fields parallel and perpendicular to the $(1,1,1)$ 
direction, respectively. Their ${\cal S}_{m,m,m}$'s 
are denoted respectively by ${\cal I}_{3m}^{\parallel}$ and 
${\cal I}_{3m}^{\perp}$ and have been computed to high orders. 
Here we only present the first few:
\begin{eqnarray*} 
{\cal I}_{3}^{\parallel}&=&6\cos\theta, \\
{\cal I}_{6}^{\parallel}&=&36+42\cos2\theta+12\cos4\theta, \\
{\cal I}_{9}^{\parallel}&=&864\cos\theta+528\cos3\theta+216\cos5\theta
+54\cos7\theta+18\cos9\theta,  \\
{\cal I}_{12}^{\parallel}&=& 7308+12504\cos2\theta+8082\cos4\theta
+4032\cos6\theta+1740\cos8\theta \\
& &\mbox{}+672\cos10\theta
+216\cos12\theta+72\cos14\theta+24\cos16\theta;  
\end{eqnarray*}
and
\begin{eqnarray*}
{\cal I}_{3}^{\perp}&=&4\cos\theta+2\cos2\theta, \\
{\cal I}_{6}^{\perp}&=&14+12\cos\theta+16\cos2\theta
+12\cos3\theta+12\cos4\theta
+8\cos5\theta+10\cos6\theta+4\cos7\theta+2\cos8\theta, \\
{\cal I}_{9}^{\perp}&=&76+204\cos\theta+176\cos2\theta+180\cos3\theta
+156\cos4\theta+156\cos5\theta+136\cos6\theta \\ 
& &\mbox{}+128\cos7\theta+102\cos8\theta+84\cos9\theta
+68\cos10\theta+64\cos11\theta+48\cos12\theta \\
& &\mbox{}+40\cos13\theta+26\cos14\theta+20\cos15\theta+10\cos16\theta
+4\cos17\theta+2\cos18\theta, \\
{\cal I}_{12}^{\perp}&=&1372+2464\cos\theta+2606\cos2\theta+2420\cos3\theta
+2502\cos4\theta+2288\cos5\theta+2288\cos6\theta+2068\cos7\theta \\ 
& &\mbox{}+2046\cos8\theta+1788\cos9\theta
+1758\cos10\theta+1532\cos11\theta+1498\cos12\theta+1264\cos13\theta \\
& &\mbox{}+1174\cos14\theta+964\cos15\theta
+894\cos16\theta+724\cos17\theta+642\cos18\theta+512\cos19\theta \\
& &\mbox{}+450\cos20\theta+340\cos21\theta
+296\cos22\theta+228\cos23\theta+178\cos24\theta+128\cos25\theta \\
& &\mbox{}+94\cos26\theta+56\cos27\theta+40\cos28\theta
+20\cos29\theta+10\cos30\theta+4\cos31\theta+2\cos32\theta.
\end{eqnarray*}
where $\theta=\phi/2$. It can be seen that ${\cal I}_{3m}^{\parallel}$ and 
${\cal I}_{3m}^{\perp}$ 
exhibit quite different behaviors as shown in Fig.~7 where we plot 
${\cal I}_{3}^{\parallel}$ through ${\cal I}_{12}^{\parallel}$ and 
${\cal I}_{3}^{\perp}$ through ${\cal I}_{12}^{\perp}$. 
Notice that the period in $\phi$ for ${\cal I}_{3m}^{\parallel}$ 
is $2\pi$ ($4\pi$) for 
even (odd) $m$, while the period for ${\cal I}_{3m}^{\perp}$ is $4\pi$ 
for any $m$. Therefore, the periodicity for the MC in 3D is identical to that 
in 2D. 

We have also computed ${\cal I}_{3m}^{\parallel}$ and 
${\cal I}_{3m}^{\perp}$ 
($m=1,2,\ldots,300$) for $\phi/2\pi=3/5$ and $(\sqrt{5}-1)/2$ and find 
that their behaviors are insensitive to the commensurability 
of $\phi$, unlike the case on a square lattice.  Physically, this 
can be understood because two randomly chosen paths have a higher 
probability of crossing (and thus interfering) in 2D than in 3D;
thus making QI effects less pronounced in 3D than in 2D.  A similar
situation occurs classically (e.g., multiply-scattered light 
in a random medium\cite{dashen,feng}). 

\subsection{Small-field magnetoconductance}
For a 3D system, the relative MC, $\Delta G(a,b,c)$, now reads
\begin{equation}
\Delta G(a,b,c)=\sqrt{1-\left[\frac{{\cal S}_{m,n,l}(2a,2b,2c)}
{{\cal N}}\right]^2}\,.
\end{equation}
The above general expression  is valid for any ending site as well as 
arbitrary orientation and strength of the magnetic field.
From Eq.~(61), in the small-field limit and at ending site 
$(r/3,r/3,r/3)$, we have 
\begin{equation}
\Delta G = \frac{1}{3\sqrt{3}}\sqrt{r^2(a^2+b^2+c^2)
+\frac{r^3}{3}\left[(b-c)^2+(c-a)^2+(a-b)^2\right]}\,,
\end{equation}
which is applicable for any orientation of the field. 
Below we focus on two special orientations of the field: 
${\bf B}_{\perp}$ and ${\bf B}_{\parallel}$.

For very small $\phi$, we have from Eq.~(61) 
\begin{equation}
\ln {\cal I}_{r}^{\perp} \simeq \ln \frac{r!}{[(r/3)!]^3}
-\frac{1}{144}\,r^{2}(r+1)\,\phi^{2}\,,
\end{equation}
and
\begin{equation}
\ln {\cal I}_{r}^{\parallel} \simeq \ln \frac{r!}{[(r/3)!]^3}
-\frac{1}{72}\,r^{2}\,\phi^{2}\,.
\end{equation}
The 3D behavior of the low-flux MC thus becomes clear: for ${\bf B}_{\perp}$
\begin{equation}
\Delta G(\phi) \simeq \frac{\sqrt{2}}{6}\,r^{3/2}\,\phi\,,
\end{equation}
and for 
${\bf B}_{\parallel}$
\begin{equation}
\Delta G(\phi) \simeq \frac{1}{3}\,r\,\phi\,.
\end{equation}  
These results can be 
interpreted as follows: the effective area $A_{\perp}^{{\rm eff}}$ 
exposed to ${\bf B}_{\perp}$ 
is larger, 
$$A_{\perp}^{{\rm eff}} \sim r^{3/2},$$ similar to the 2D case where 
$\Delta G(\phi) \propto r^{3/2}\,\phi$; while the effective area 
$A_{\parallel}^{{\rm eff}}$ exposed to 
${\bf B}_{\parallel}$ is smaller, 
$$A_{\parallel}^{{\rm eff}} \sim r,$$ 
thus closer to our quasi-1D ladder 
case with $\Delta G(\phi) \propto r\,\phi$. 

As a numerical test of Eqs.~(66) 
and (67), in Fig.~8 we show $\Delta G$ calculated directly from Eq.~(62), 
versus $r^{3/2}\phi$ in (a) and versus $r\phi$ in (b), respectively for 
several values of ${\bf B}_{\perp}$ and ${\bf B}_{\parallel}$ with hopping 
length $r=3, 6,\ldots, 600$. The collapse of all the data into a straight 
line verifies the scaling of the low-flux $\Delta G$ 
presented above. 

\subsection{Average of the magnetoconductance over angles}
In a macroscopic sample, the conductance may be determined by {\em a few} 
(instead of one, as considered before) critical hopping events. 
As a result of this, the observed MC of the 
whole sample should be the average of the MC associated with 
these critical hops. Thus, in 3D systems 
it is also important to take into account 
the randomness of the angles between the hopping direction and the 
orientation of the applied magnetic field.\cite{AM} 

To theoretically investigate the 
effect of the average over angles on the MC, we consider all possible 
relative hopping directions with respect to that of the magnetic field, 
or equivalently, the continuously-varing orientation of the field 
with respect to a fixed hopping direction. 
We adopt the latter below: the ending site of all hopping 
events (with the same hopping length $r$) is located at the 
diagonal point $(r/3,r/3,r/3)$ and the magnetic field can be adjusted 
between the parallel and perpendicular directions with respect to 
the vector ${\bf d}=(1,1,1)$. Our interest here is in 
the MC averaged over angles, 
denoted by $\overline{\Delta G}$, in the low-field limit. 
Recall that the magnetic field is ${\bf B}=(a,b,c)/2\pi$, and from Eq.~(63), 
we have
\begin{equation}
\Delta G =\frac{2\pi}{3\sqrt{3}}\,r\,B\,\sqrt{1+r\sin^2\omega}\,,
\end{equation}
where $B=\sqrt{a^2+b^2+c^2}/2\pi$ is the magnitude of the field and 
$\omega$ is the angle between ${\bf B}$ and ${\bf d}$.
By averaging over the angle $\omega$, we obtain
\begin{eqnarray}
\overline{\Delta G}(B) &=&\frac{4}{3\sqrt{3}}\,r\,B\int_{0}^{\pi/2} 
\sqrt{1+r\sin^2\omega}\,d\omega  \nonumber \\ 
&=&\frac{4}{3\sqrt{3}}\,r\sqrt{r+1}\,B\,
E\!\left(\frac{\pi}{2},\frac{\sqrt{r}}{\sqrt{r+1}}\right)\,,
\end{eqnarray}
where $E(\pi/2,\sqrt{r}/\sqrt{r+1})$ is the complete elliptic 
integral of the second kind. 
When $r$ is large, $E(\pi/2,\sqrt{r}/\sqrt{r+1})\simeq 1$ and 
we therefore have 
\begin{equation}
\overline{\Delta G}(B)\simeq\frac{4}{3\sqrt{3}}\,r^{3/2}\,B.
\end{equation}
Equation~(70) means that the dominant contribution to the MC stems from the 
critical hop which is {\em perpendicular} to the field. This is understandable 
through our earlier observation that the effective area enclosed by the 
electron is largest when ${\bf B}$ is perpendicular to ${\bf d}$.
From the above analysis, we conclude that in 3D macroscopic samples 
the low-field MC should in principle behave as $r^{3/2}B$.

\section{Concluding remarks and summary of results}
In closing we briefly address four issues.  First, 
although relevant measurable quantities such as $|S_{m,n}|^{2}$ and 
$|{\cal S}_{m,n,l}|^{2}$ are gauge-invariant, the transition amplitudes 
are gauge dependent.  As an illustration, the transition amplitude will be 
$L_{m,n}$ [Eq.~(12)] if we use the
Landau gauge ${\bf A}=(0,Bx)$ on a square lattice. 
The notation $S_{m,n}$ ($L_{m,n}$) refers the use of 
the $S$ymmetric ($L$andau) gauge. Similarly, the transition amplitude will 
read ${\cal L}_{m,n,l}$ [Eq.~(58)] if we use the 
gauge ${\bf A}=(B_yz,B_zx,B_xy)$ on a cubic lattice.

Second, returns to the origin (see, e.g., Refs.~1 and 22-24) become 
important for less strongly localized electrons, and their 
QI effects\cite{lin,nori} 
can be incorporated 
in our approach.  

Third, the main limitations of our study in the case 
with impurities are the 
following: no inclusion of
spin-orbit scattering effects (for this see, e.g., Refs.~7-9 
and references therein), and no explicit inclusion of the correlations 
between crossing paths, as discussed in Refs.~4 and 7. However, 
these correlations are negligible when spin-orbit scattering is 
present.\cite{6} 

Fourth, besides analytical closed-form results in 2D, this work presents 
exact results for 3D systems, e.g., 
$\Delta G=(2\pi/3\sqrt{3})rB(1+r\sin^2\omega)^{1/2}$ [Eq. ~(68)].
These results can provide further tests of the quantum interference effects.  
This can be done by measuring the MC of bulk samples 
(which are small enough that only  a single critical hop is allowed) 
in various orientations of the field.  By doing this, one can then 
{\it determine\/} the values of  $r$ and $\omega$ (and, hence, also 
the direction of this critical hop). 
Therefore, the small-field behaviors of the MC with fields parallel and 
perpendicular to the direction of this critical hop can be measured 
and compared to our predictions.  This could potentially be very useful.

In summary, we present an investigation of quantum interference phenomena
and the magnetic-field effects on the MC 
resulting from sums over directed paths on 
resistor networks in 2D and 3D. The principal results include: (1) an 
exact and explicit closed-form expression for the sum over 
forward-scattering paths $S^{(r)}$ to any site on a square lattice, 
which is the essential QI quantity in both uniform and disordered cases, 
(2) an explicit formula for $S^{(r)}$ for electrons hopping on a cubic 
lattice, (3) the low-flux 
behaviors of $S^{(r)}$ in both 2D and 3D, (4) the exact summation of the 
dominant winding paths in 2D, (5) compact, analytic results for the 
positive MC as explicit functions of the magnetic flux 
which are valid in any dimension, (6) the small-field behaviors of 
the MC in quasi-1D, 2D and 3D, and (7) an analytic result for the small-field 
MC in 3D incorporating the randomness in the relative angles between the 
hops and the applied field. 
They provide analytical and explicit closed-form 
results concerning the hopping transport of strongly localized electrons 
subject to an external magnetic field in the macroscopic regime. 
We hope that our results stimulate further work 
(e.g., inclusion of spin-orbit effects on lattice path integrals) 
on exact results in 2D and 3D systems.

\acknowledgments

It is a great pleasure to thank B.L. Altshuler, M. Kardar, E. Medina, 
Y. Meir, Y. Shapir and 
J. Stembridge for their very useful suggestions. 
F.N. acknowledges partial support from 
the UCSB Institute for Theoretical 
Physics through the NSF grant No. PHY89-04035, 
the NATO office of Scientific Affairs, and the Office of the Vice Provost
for Academic and Multicultural Affairs, through a Faculty Career Development
Award during the Fall of 1994.

\appendix
\section{Derivation of the moments $\langle \,J^{2p}_{m,n}(\phi)\, \rangle$ 
and $\langle \,|J_{m,n}(\phi)|^{2p}\, \rangle$}

In this appendix, we outline the derivation of the moments shown in 
Eqs.~(32), (33), (35), and (36). 
First, let us derive Eqs.~(32) and (33). Let $s_i =\exp(i\Phi_{i})$, 
where $\Phi_i$ is the sum over all phases along path $i$ and 
$i=1, 2, \ldots, N$. Note that for {\em every} path, the probability 
for obtaining the overall phase factor $\pm s_i$ is $P_{\pm}$. Now
$$J^2=\left(\sum_{i=1}^{N}\gamma_i s_i\right)\,
\left(\sum_{i=1}^{N}\gamma_i s_i\right)$$
and
$$|J|^2=\left(\sum_{i=1}^{N}\gamma_i s_i\right)\,
\left(\sum_{i=1}^{N}\gamma_i \frac{1}{s_i}\right),$$
where $\gamma_i=\pm 1$ with probability $P_{\pm}$. 
If the number of $\gamma_i=-1$ is $k$, the overall probability is 
$P_{+}^{N-k}\,P_{-}^{k}$ and there are $C_k^N$ combinations among 
$\gamma_{i}$ ($i=1, \cdots, N$). For $k=0$ and $k=N$, there is only one 
combination producing $S(2\phi)+2\sum_{i\neq j}s_i\,s_j$ 
for $J^2$  and also only one 
combination producing $N+\sum_{i\neq j} s_i/s_j$ for $|J|^2$. 
When $1 \leq k \leq N-1$, for $J^2$ there are $N_{-}=2\,C_{k-1}^{N-2}$ 
combinations producing $S(2\phi)-2\sum_{i\neq j}s_i\,s_j$ and 
$N_{+}=C_k^N-2\,C_{k-1}^{N-2}$ combinations producing 
$S(2\phi)+2\sum_{i\neq j}s_i\,s_j$. Also, when $1 \leq k \leq N-1$, 
for $|J|^2$ there are $N_{-}$ combinations 
producing $N-\sum_{i\neq j}s_i/s_j$ and $N_{+}$ combinations producing 
$N+\sum_{i\neq j}s_i/s_j$. 
In $N_{-}$, the factor $2$ comes from the two possible 
ways: $+s_i-s_j$ and $-s_i+s_j$. $C_{k-1}^{N-2}$ comes from arranging 
$(k-1)$'s minus signs among the 
$(N-2)$\ $s_l$'s left ($l \neq i \neq j$). Therefore, the overall average 
is ${\cal P}=P_{+}^N+P_{-}^N
+\sum_{k=1}^{N-1}(N_{+}-N_{-})\,P_{+}^{N-k}\,P_{-}^{k}
=(\mu-\nu)^{2r}.$  We thus have
$$\langle \,J^2\,\rangle =S(2\phi)
+{\cal P}\,\left(2\sum_{i\neq j}s_i\,s_j\right),$$
where we have used the relation 
$\sum_{k=0}^{N}\,C_{k}^{N}\,P_{+}^{N-k}\,P_{-}^{k}=(P_{+}+P_{-})^N=1.$ 
Similarly,
$$\langle \,|J|^2\,\rangle=N+{\cal P}\,\sum_{i\neq j}\frac{s_i}{s_j}\,.$$ 
By exploiting
$$2\sum_{i\neq j}s_i\,s_j =S^2(\phi)-S(2\phi)$$ and 
$$\sum_{i\neq j}\frac{s_i}{s_j} =S^2(\phi)-N\,,$$
we thus obtain Eqs.~(32) and (33).

For $\mu=\nu=1/2$, (namely, each site contributes either $+1$ or $-1$ 
with equal probability), $P_{\pm} = 1/2$ for {\em every} path $i$. 
The total number of sites that can be visited is $T=(m+1)(n+1)-1$ 
(the initial site is not counted). Therefore, the total number of all 
possible impurity configurations is $2^{T}$. 
Let us now focus on the $\gamma_i\,s_i$ with $i=1, 2, \cdots , N.$ 
The total number of configuration sets is $2^{N}$ since each $\gamma_i$ can 
be either $+1$ or $-1$ with equal probability. Among the $2^{T}$ 
impurity configurations, there are {\em always} $2^T/2^N$ producing 
a set of $\gamma_i s_i$ for {\em every} 
possible set of $\gamma_i s_i$. For instance, $T=3$ and $N=2$ in the 
simplest case $S_{1,1}$. Among $2^3=8$ impurity configurations, two of 
them give $+s_1+s_2$, $+s_1-s_2$, $-s_1+s_2$, and $-s_1-s_2$. 
We thus have
$$
\langle \, J^{2p}\,\rangle =\frac{1}{2^N}\,\sum_{\{\gamma_i\}}
\left[\left(\sum_{i=1}^{N} \gamma_i\,s_i\right)^{p}\left(\sum_{i=1}^{N} 
\gamma_i\,s_i\right)^{p}\right],
$$
and similarly,
$$\langle \, |J|^{2p}\,\rangle=\frac{1}{2^N}\,\sum_{\{\gamma_i\}}
\left[\left(\sum_{i=1}^{N} \gamma_i\,s_i\right)^{p}\left(\sum_{i=1}^{N} 
\gamma_i\,\frac{1}{s_i}\right)^{p}\right].$$
The summation in $\{\gamma_i\}$ is over all possible 
configuration sets of $\gamma_i$. 
In other words, the average of $A$ over disorder 
(i.e., $\langle \, A\,\rangle$) means to sum over all possible 
$\gamma_i =\pm 1$ for the desired quantity $A$, and then divide the sum 
by $2^{N}$. We thus obtain
\begin{eqnarray*}
\langle \, J^4\,\rangle&=&S(4\phi)\,+\,\frac{4!}{(2!)^2}
\sum_{i\neq j}s_i^2\,s_j^2\,, \\
\langle \, J^6\,\rangle&=&S(6\phi)\,+\,\frac{6!}{4!\,2!}
\sum_{i\neq j}s_i^4\,s_j^2
\,+\,\frac{6!}{(2!)^3}\sum_{i\neq j\neq k}s_i^2\,s_j^2\,s_k^2\,, \\
\langle \, J^8\,\rangle&=&S(8\phi)\,+\,\frac{8!}{6!\,2!}
\sum_{i\neq j}s_i^6\,s_j^2
\,+\,\frac{8!}{(4!)^2}\sum_{i\neq j}s_i^4\,s_j^4
\,+\,\frac{8!}{4!\,(2!)^2}\sum_{i\neq j\neq k}s_i^4\,s_j^2\,s_k^2
\,+\,\frac{8!}{(2!)^4}
\sum_{i\neq j\neq k\neq l}s_i^2\,s_j^2\,s_k^2\,s_l^2\,, \\
\langle \, J^{10}\,\rangle&=&S(10\phi)\,+\,\frac{10!}{8!\,2!}
\sum_{i\neq j}s_i^8\,s_j^2
\,+\,\frac{10!}{6!\,4!}\sum_{i\neq j}s_i^6\,s_j^4
\,+\,\frac{10!}{6!\,(2!)^2}\sum_{i\neq j\neq k}s_i^6\,s_j^2\,s_k^2
\,+\,\frac{10!}{(4!)^2\,2!}\sum_{i\neq j\neq k}s_i^4\,s_j^4\,s_k^2 \\
& & \mbox{}+\frac{10!}{4!\,(2!)^3}\sum_{i\neq j\neq k\neq l}
s_i^4\,s_j^2\,s_k^2\,s_l^2
\,+\,\frac{10!}{(2!)^5}\sum_{i\neq j\neq k\neq l\neq m}
s_i^2\,s_j^2\,s_k^2\,s_l^2\,s_m^2\,;
\end{eqnarray*}
and
\begin{eqnarray*}
\langle \, |J|^4\,\rangle&=&N\,(2N-1)\,+\,\sum_{i\neq j}
\frac{s_i^2}{s_j^2}\,, \\
\langle \, |J|^6\,\rangle&=&N\,(6N^2-9N+4)\,+\,3\,(3N-4)
\sum_{i\neq j}\frac{s_i^2}{s_j^2}\,, \\
\langle\,|J|^8\,\rangle&=&N\,(24N^3-72N^2+82N-33)\,
+\,4\,(18N^2-57N+49)\sum_{i\neq j}\frac{s_i^2}{s_j^2}\,
+\,\sum_{i\neq j}\frac{s_i^4}{s_j^4} \\
& & \mbox{}+6\sum_{i\neq j\neq k}\left(\frac{s_i^4}{s_j^2\,s_k^2}
+\frac{s_j^2\,s_k^2}{s_i^4}\right)\,+\,\sum_{i\neq j\neq k\neq l}
\frac{s_k^2\,s_l^2}{s_i^2\,s_j^2}\,, \\
\langle\,|J|^{10}\,\rangle&=&N\,(120N^4-600N^3+1250N^2-1225N+456)\,
+\,20\,(30N^3-165N^2+325N-224)\sum_{i\neq j}\frac{s_i^2}{s_j^2} \\
& & \mbox{}+5\,(5N-8)\sum_{i\neq j}\frac{s_i^4}{s_j^4}\,+\,10\,(15N-32)
\sum_{i\neq j\neq k}\left(\frac{s_i^4}{s_j^2\,s_k^2}+\frac{s_j^2\,s_k^2}
{s_i^4}\right)
\,+\,300\,(3N-8)\sum_{i\neq j\neq k\neq l}\frac{s_i^2\,s_j^2}{s_k^2\,s_l^2}\,. 
\end{eqnarray*}
Noticing that $\sum_{i=1}^{N}s_i^m=\sum_{i=1}^{N}1/s_i^m=S(m\phi)$, 
it can be derived that:
\begin{eqnarray*}
\sum_{i\neq j}s_i^2\,s_j^2 &=&\frac{1}{2}[S^2(2\phi)-S(4\phi)], \\
\sum_{i\neq j}s_i^4\,s_j^2 &=&S(2\phi)\,S(4\phi)-S(6\phi), \\
\sum_{i\neq j}s_i^4\,s_j^4 &=&\frac{1}{2}[S^2(4\phi)-S(8\phi)], \\
\sum_{i\neq j}s_i^6\,s_j^2 &=&S(2\phi)\,S(6\phi)-S(8\phi), \\
\sum_{i\neq j}s_i^8\,s_j^2 &=&S(2\phi)\,S(8\phi)-S(10\phi), \\
\sum_{i\neq j}s_i^6\,s_j^4 &=&S(4\phi)\,S(6\phi)-S(10\phi), \\
\sum_{i\neq j\neq k}s_i^2\,s_j^2\,s_k^2 &=& 
\frac{1}{6}\,S^3(2\phi)-\frac{1}{2}\,S(2\phi)\,S(4\phi)
+\frac{1}{3}\,S(6\phi),  \\
\sum_{i\neq j\neq k}s_i^4\,s_j^2\,s_k^2 &=& 
\frac{1}{2}\,S^2(2\phi)\,S(4\phi)-S(2\phi)\,S(6\phi)
-\frac{1}{2}\,S^2(4\phi)+S(8\phi), \\
\sum_{i\neq j\neq k}s_i^6\,s_j^2\,s_k^2 &=&\frac{1}{2}\,S^{2}(2\phi)
\,S(6\phi)-S(2\phi)\,S(8\phi)
-\frac{1}{2}\,S(4\phi)\,S(6\phi)+S(10\phi), \\
\sum_{i\neq j\neq k}s_i^4\,s_j^4\,s_k^2 &=&
\frac{1}{2}\,S(2\phi)\,S^{2}(4\phi)-\frac{1}{2}\,S(2\phi)\,S(8\phi)
-S(4\phi)\,S(6\phi)+S(10\phi),  \\
\sum_{i\neq j\neq k\neq l}s_i^2\,s_j^2\,s_k^2\,s_l^2 &=& 
\frac{1}{24}\,S^4(2\phi)-\frac{1}{4}\,S^2(2\phi)\,S(4\phi)
+\frac{1}{3}\,S(2\phi)\,S(6\phi)+\frac{1}{8}\,S^2(4\phi)
-\frac{1}{4}\,S(8\phi),\\
\sum_{i\neq j\neq k\neq l}s_i^4\,s_j^2\,s_k^2\,s_l^2 &=&
 \frac{1}{6}\,S^{3}(2\phi)\,S(4\phi)-\frac{1}{2}\,S^2(2\phi)\,S(6\phi)
-\frac{1}{2}\,S(2\phi)\,S^2(4\phi), \\ 
& & \mbox+S(2\phi)\,S(8\phi)+\frac{5}{6}\,S(4\phi)\,S(6\phi)-S(10\phi),  \\
\sum_{i\neq j\neq k\neq l\neq m}s_i^2\,s_j^2\,s_k^2\,s_l^2\,s_m^2 &=&
\frac{1}{120}\,S^5(2\phi)-\frac{1}{12}\,S^3(2\phi)\,S(4\phi)
+\frac{1}{6}\,S^2(2\phi)\,S(6\phi)+\frac{1}{8}\,S(2\phi)\,S^2(4\phi)  \\
& &\mbox{}-\frac{1}{4}\,S(2\phi)\,S(8\phi)-\frac{1}{6}\,S(4\phi)\,S(6\phi)
+\frac{1}{5}\,S(10\phi),  \\
\sum_{i\neq j}\frac{s_i^2}{s_j^2} &=& S^2(2\phi)-N, \\
\sum_{i\neq j}\frac{s_i^4}{s_j^4}&=& S^2(4\phi)-N, \\
\sum_{i\neq j\neq k}\frac{s_i^4}{s_j^2\,s_k^2}&=&
\sum_{i\neq j\neq k}\frac{s_j^2\,s_k^2}{s_i^4}
=\frac{1}{2}S^2(2\phi)\,S(4\phi)-S^2(2\phi)-\frac{1}{2}S^2(4\phi)+N, \\
\sum_{i\neq j\neq k\neq l}\frac{s_i^2\,s_j^2}{s_k^2\,s_l^2} &=& 
\frac{1}{4}\,S^4(2\phi)-(N-2)\,S^2(2\phi)-\frac{1}{2}\,S^2(2\phi)\,S(4\phi)
+\frac{1}{4}\,S^2(4\phi)
+\frac{1}{2}N\,(N-3).
\end{eqnarray*}
By utilizing the above relations, we have obtained the results for the 
moments presented in Eqs.~(35.1)-(35.4) and (36.1)-(36.4).

\section{A different model of disorder}
In this appendix, we study another model of diagonal disorder: 
$\epsilon_{i}$ uniformly distributed between $-W/2$ and $W/2$. 
Our focus is on the analytical computation of the leading terms 
(terms $\propto N^p$) in the moments $\langle\,|J(0)|^{2p}\,\rangle$ and 
$\langle\,|J(\phi)|^{2p}\,\rangle$. Indeed, the scheme presented 
below applies equally  to our first model of disorder, where $\epsilon_i$ 
can take two values, $+W$ and $-W$, with equal probability. 

Let us write 
$$J(\phi)=\sum_{i=1}^{N}\eta_{i}s_i\,,$$
where for each path $i$ 
$$\eta_i=\prod_{j=1}^{r}\left(-\frac{W}{\epsilon_j}\right).$$
It is reasonable to choose $\eta_i$  from a Gaussian distribution 
of zero mean and unit standard deviation. We therefore have 
$\langle \eta_i^{2n}\rangle=1$ and $\langle \eta_i^{2n+1}\rangle=0$.
For $\phi=0$, we have
$$|J(0)|^{2p} = \left(\sum_{i=1}^{N}\eta_i\right)^{2p}=
\prod_{k=1}^{2p}\left(\sum_{i_k=1}^{N} \eta_{i_k}\right)\,.$$
It is found that the leading term in $\langle \,|J(0)|^{2p}\,\rangle$ 
comes from all terms 
having the form $\prod_{i=1}^{p}\eta_{i}^2$ and there are 
$C^{N}_{p}$ distinct terms of this type, each term has a coefficient 
$(2p)!/2^p$. Therefore, we obtain for the 
leading term
$$\langle \,|J(0)|^{2p}\,\rangle 
=\frac{(2p)!}{2^p}\,\frac{N^p}{p!}=(2p-1)!!\,N^p.$$

For $\phi\neq 0$, we have now
\begin{eqnarray*}
|J(\phi)|^{2p}& = & \left(\sum_{i=1}^{N}\eta_is_i\right)^{p}
\left(\sum_{i=1}^{N}\eta_i\frac{1}{s_i}\right)^{p} \\
&=&\prod_{k=1}^{p}\left[\left(\sum_{i_k=1}^{N} \eta_{i_k}s_{i_k}\right)
\left(\sum_{i^{\prime}_{k}=1}^{N} \eta_{i^{\prime}_{k}}
\frac{1}{s_{i^{\prime}_{k}}}\right)\right]\,.
\end{eqnarray*}
The contributions to the leading terms involve different factors, 
as shown below. 
There are $C^{N}_{p}$ terms like 
$$\left(\prod_{i=1}^{p}\eta_i s_i\right)
\left(\prod_{i=1}^{p}\eta_i\frac{1}{s_i}\right),$$ 
and they contribute
$$(p!)^{2}\ \frac{N^p}{p!}=p!\,N^p.$$
There are $C^{N-2}_{p-2}$ terms like
$$\left(\eta_1^2s_1^2\prod_{i=2}^{p-1}\eta_i s_i\right)
\left(\eta_{1'}^2\frac{1}{s_{1'}^2}
\prod_{i=2}^{p-1}\eta_i\frac{1}{s_i}\right),$$
and they contribute
$$\left(\frac{p!}{2}\right)^{2}
\,\frac{N^{p-2}}{(p-2)!}\,\frac{S^2(2\phi)}{(1!)^2}
=p!\,\frac{2!\,C^p_2}{(2\cdot 1!)^2}\,N^{p-2}\,S^2(2\phi).$$ 
Similarly, there are $C^{N-4}_{p-4}$ terms like
$$\left(\eta_1^2\,s_1^2\,\eta_2^2\,s_2^2\prod_{i=3}^{p-2}\eta_i\,s_i\right)
\left(\eta_{1'}^2\,\frac{1}{s_{1'}^2}\,\eta_{2'}^2\,\frac{1}{s_{2'}^2}
\prod_{i=3}^{p-2}\eta_i\,\frac{1}{s_i}\right),$$
and they contribute
$$\left(\frac{p!}{2^2}\right)^{2}
\,\frac{N^{p-4}}{(p-4)!}\,\frac{S^4(2\phi)}{(2!)^2}
=p!\,\frac{4!\,C^p_4}{(2^2\cdot 2!)^2}\,N^{p-4}\,S^4(2\phi).$$
In general, there are $C^{N-2k}_{p-2k}$ terms of the following type
$$\left(\prod_{i=1}^{k}\eta_i^2 s_i^2
\prod_{i=k+1}^{p-k}\eta_i s_i\right)
\left(\prod_{i=1}^{k}\eta_{i'}^2\frac{1}{s^2_{i'}}
\prod_{i=k+1}^{p-k}\eta_i\frac{1}{s_i}\right),$$
each one with a coefficient $(p!/2^k)^2$. The contribution to the moment is 
therefore 
$$\left(\frac{p!}{2^{k}}\right)^2\,\frac{N^{p-2k}}{(p-2k)!}
\,\frac{S^{2k}(2\phi)}{(k!)^2}
=p!\,\frac{(2k)!\,C^p_{2k}}{(2^{k}\,k!)^2}N^{p-2k}\,S^{2k}(2\phi).$$
Notice that we have utilized the fact that
$$\sum_{{\rm All\ different}\ j_i\ {\rm and}\ l_i}
\left(\prod_{i=1}^{k}\frac{s_{j_i}^2}{s_{l_i}^2}\right)$$ 
always contains $S^{2k}(2\phi)/(k!)^2$. 
Totally, we thus obtain for the leading terms
$$\langle\, |J(\phi)|^{2p} \,\rangle = p!\,N^{p}\,\left\{\sum_{k=0}^{\infty}
\frac{(2k)!\,C^{p}_{2k}}{(2^k\,k!)^{2}}\,
\left[\frac{S(2\phi)}{N}\right]^{2k}\right\}.$$

\newpage
{\bf FIGURE CAPTIONS}

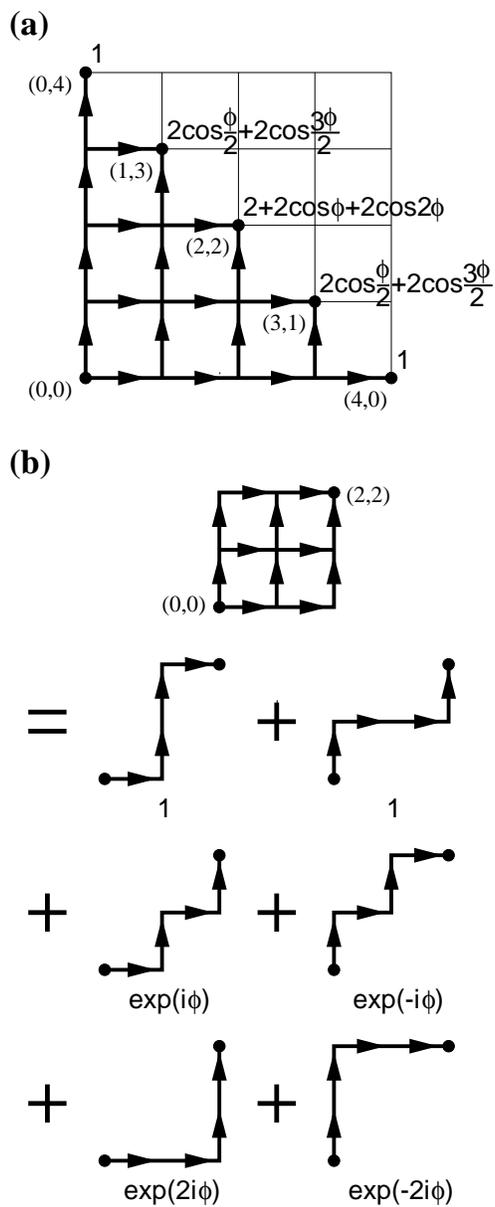
\begin{figure}
\caption{(a) Starting from $(0,0)$ on a square lattice, for 
forward-scattering paths of four steps, electrons can end at five sites: 
$(4,0)$, $(3,1)$, $(2,2)$, $(1,3)$, and $(0,4)$. 
Their corresponding $S_{m,n}$ are also shown. The arrows specify the 
electron hopping directions (only moving to the right and upward are 
allowed in the directed-path model). Notice that the symmetry 
$S_{m,n}=S_{n,m}$ holds. $S_{2,2}$ has the strongest interference 
among them because the number of paths ending at $(2,2)$, and the 
area they enclose, are both the largest. (b) Six different directed 
paths connecting $(0,0)$ and $(2,2)$ and their separate phase-factor 
contributions to $S_{2,2}$; the total equals $1+1+e^{i\phi}+e^{-i\phi}
+e^{2i\phi}+e^{-2i\phi}=2+2\cos\phi+2\cos2\phi$.}
\label{fig1}
\end{figure}

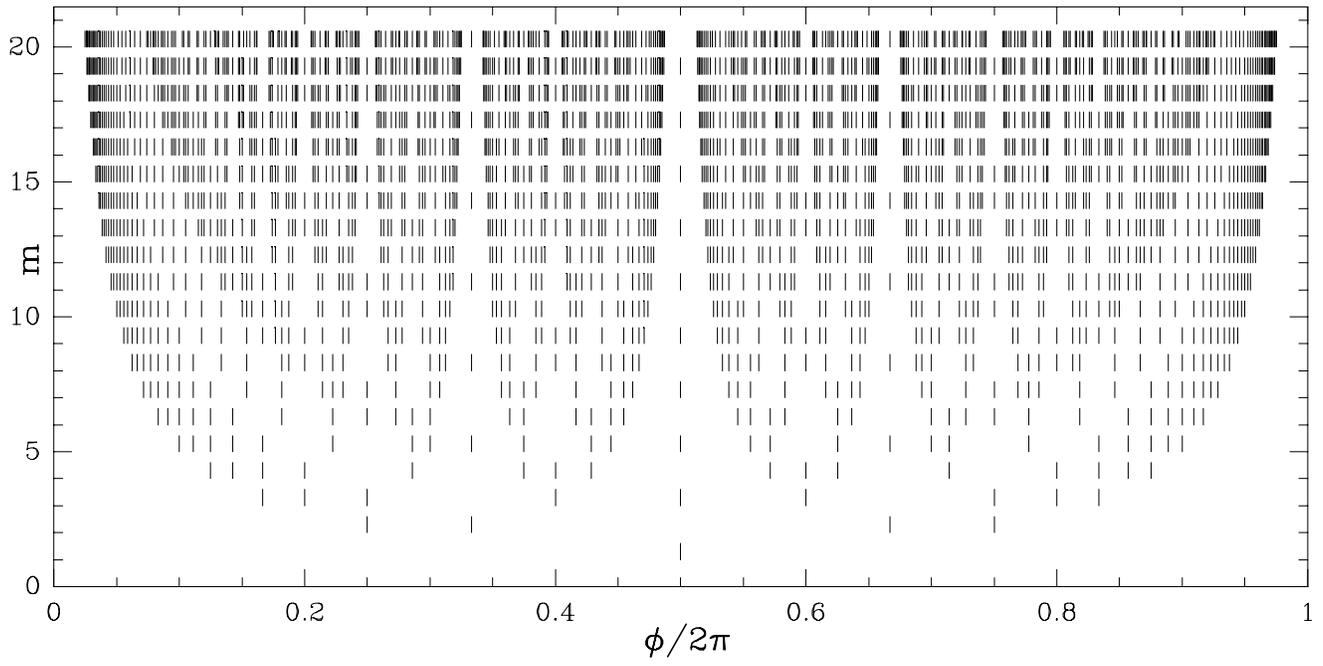
\begin{figure}
\caption{Plot of $m$ versus $\phi/2\pi$ (denoted by short bars in order 
to visualize them better), between $0$ and $1$, such that $I_{2m}(\phi)=0$; 
for $m=1, 2, \ldots, 20$. Note that the smallest one is always $1/2m$ and the 
number of zeros increases rapidly when $m$ becomes larger. The properties of 
$I_{2m}$ described in Eqs.~(19-22) are exhibited in the figure. 
For instance, when $\phi/2\pi=1/5$, $I_{2m}=0$ for $m=3+5n$ and $m=4+5n$ with 
$n \geq 0$ (namely, $m=3, 4, 8, 9, 13, 14, 18, 19,\ldots$).}
\label{fig2}\end{figure}

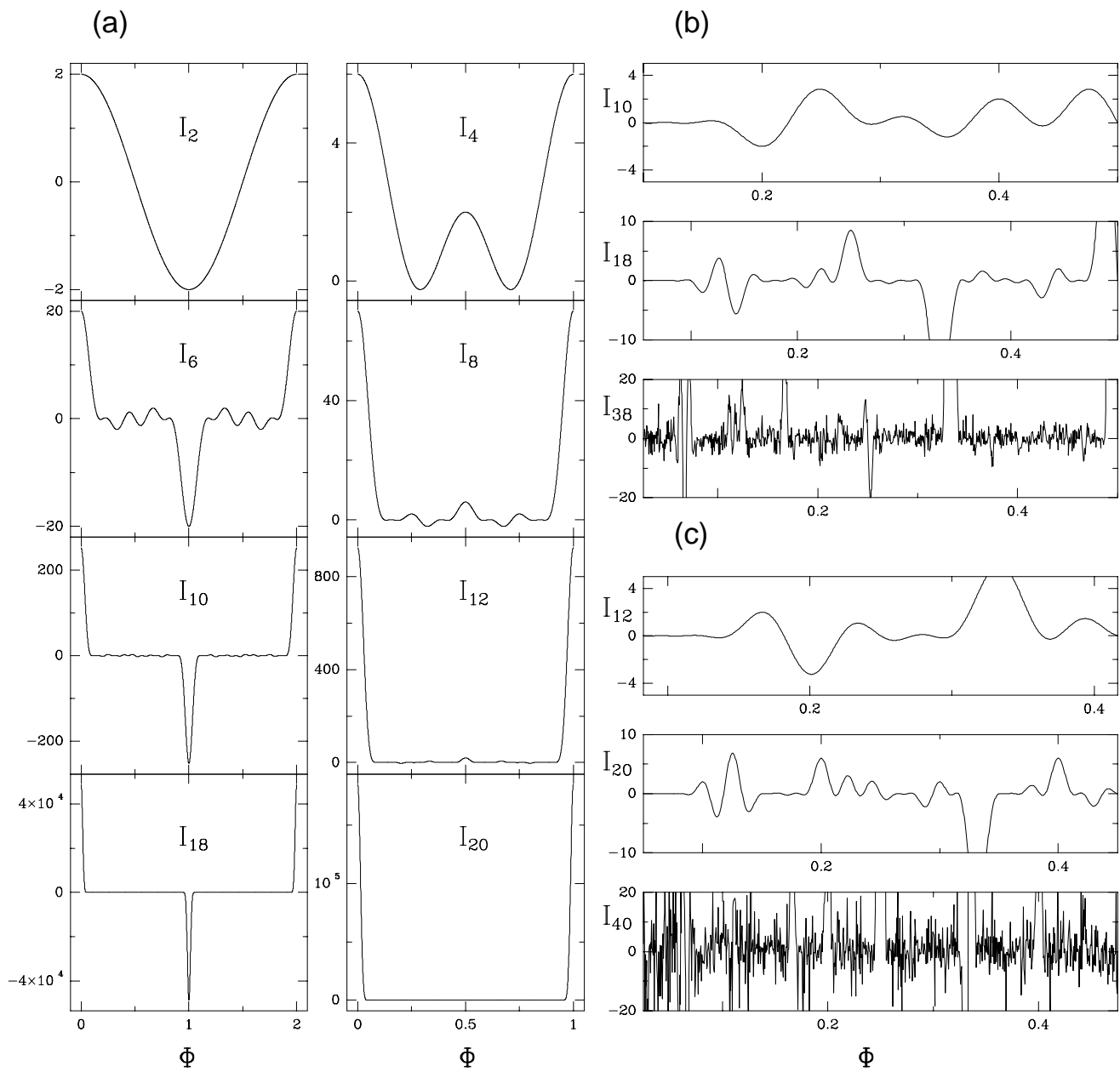
\begin{figure}
\caption{$I_{2m}$ for various $m$ 
as functions of the flux through each elementary 
plaquette, $\Phi=\phi/2\pi$ in their respective full period. 
Notice the $2\pi$ $(4\pi)$ periodicity in $\phi$ for even (odd) $m$. 
In (a), we plot $I_2, I_{4}, \ldots, I_{12}$, 
$I_{18}$, and $I_{20}$. To show the behavior of the rapid small-magnitude 
fluctuations around zero of $I_{2m}(\Phi)$ for 
$\frac{1}{2m}\leq \Phi \leq \frac{1}{2}$ when $m$ is even and for 
$\frac{1}{2m} \leq \Phi \leq \frac{m-1}{2m}$ when $m$ is odd: 
In (b), we plot $I_{10}$ (top), $I_{18}$ (middle), 
and $I_{38}$ (bottom) for $\Phi$ in their respective interval 
$[\frac{1}{2m},\frac{1}{2}]$. In (c), we plot $I_{12}$ (top), $I_{20}$ 
(middle), and $I_{40}$ (bottom) for $\Phi$ in their respective interval 
$[\frac{1}{2m},\frac{m-1}{2m}]$. Only some restricted ranges in the 
vertical axes are exhibited. From these figures, we 
clearly see the general properties for the behavior of $I_{2m}$ 
described in Sec.~II C.}
\label{fig3}
\end{figure}

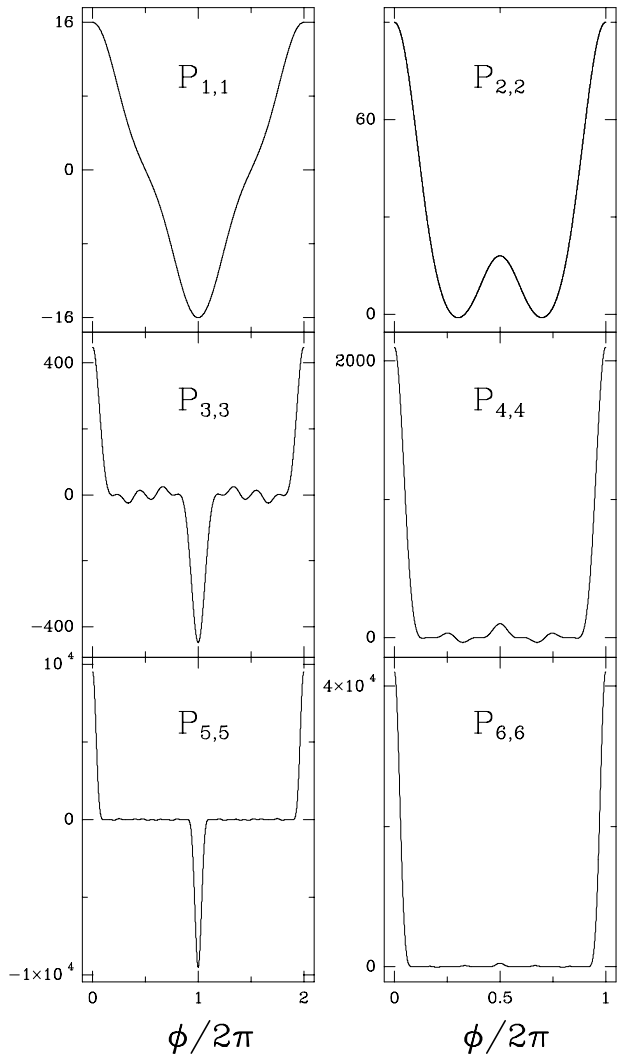
\begin{figure}
\caption{$P_{m,m}$ (for $m=1, 2,\ldots, 6$) as functions of the 
flux through each elementary plaquette, $\Phi=\phi/2\pi$.}
\label{fig4}
\end{figure}

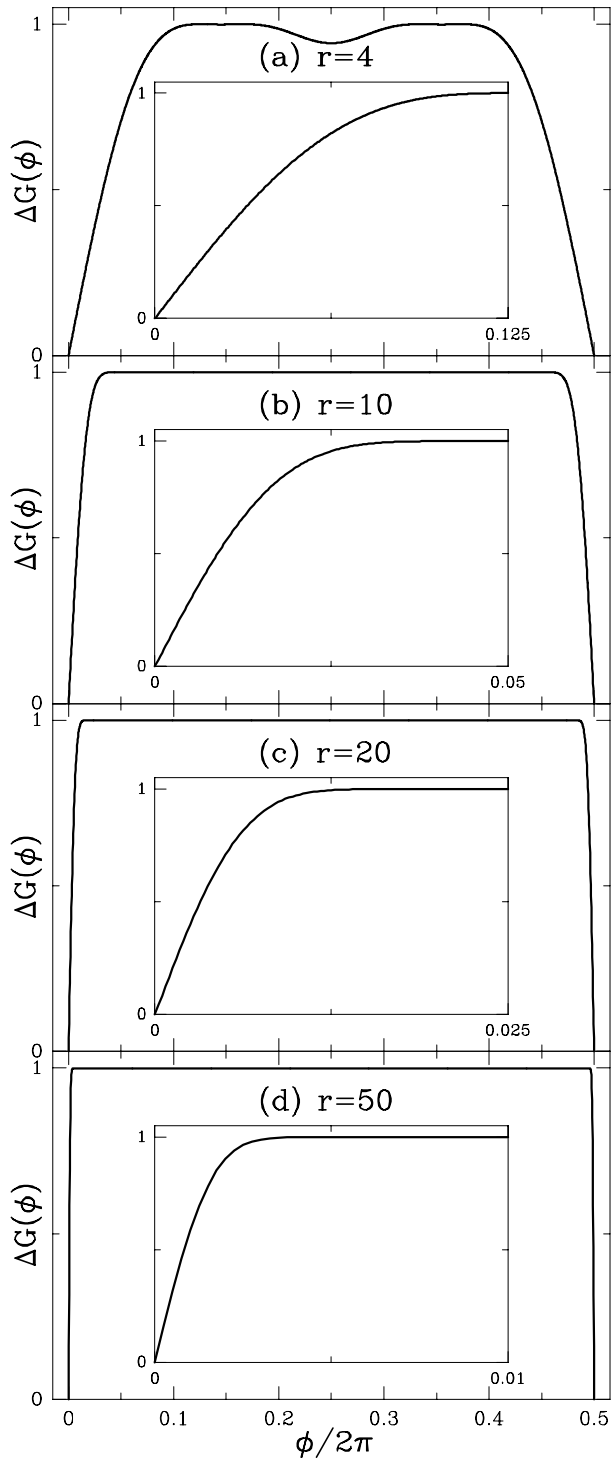
\begin{figure}
\caption{The relative magnetoconductance $\Delta G(\phi)$ versus $\phi/2\pi$ 
for hopping between $(0,0)$ and $(r/2,r/2)$ for several system sizes. 
From (a) to (d), the hopping length $r$ corresponds to $4$, $10$, $20$, and 
$50$, respectively. Inserts show $\Delta G(\phi)$ for $\phi$ between $0$ and 
the corresponding saturated field $\phi/2\pi=1/2r$. It is observed that for 
large systems (i.e., $r$ large), $\Delta G(\phi)$ rapidly approaches the 
saturation value $1$ even at $\phi/2\pi$, which is less than $1/2r$.}
\label{fig5}
\end{figure}

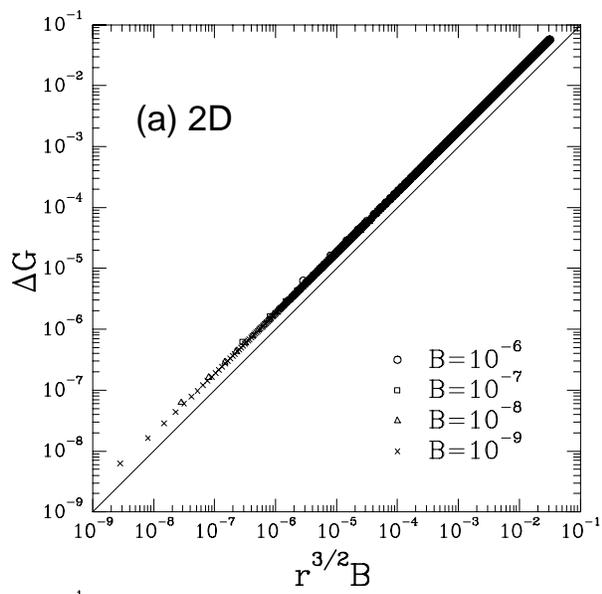
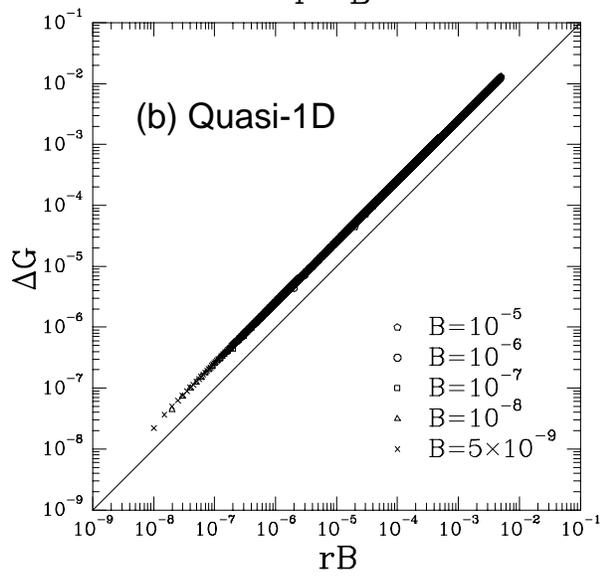
\begin{figure}
\caption{(a) $\Delta G$ versus $r^{3/2}B$ in 2D with 
the hopping length $r=2, 4,\ldots, 1000$, and (b) 
$\Delta G$ versus $rB$ in quasi-1D with 
$r=2, 3,\ldots, 500$, for various small values 
of $B$. All the data nicely collapse into a 
straight line, which verifies the scaling behavior of the small-field 
$\Delta G$: $(\protect\sqrt{3}/6)r^{3/2}\phi$ in 2D and 
$(\protect\sqrt{3}/3)r\phi$ for quasi-1D systems. 
The distance between these data 
and the solid reference line reflects the prefactor: 
$\protect\sqrt{3}/6$ in (a) and $\protect\sqrt{3}/3$ in (b).}
\label{fig6}
\end{figure}

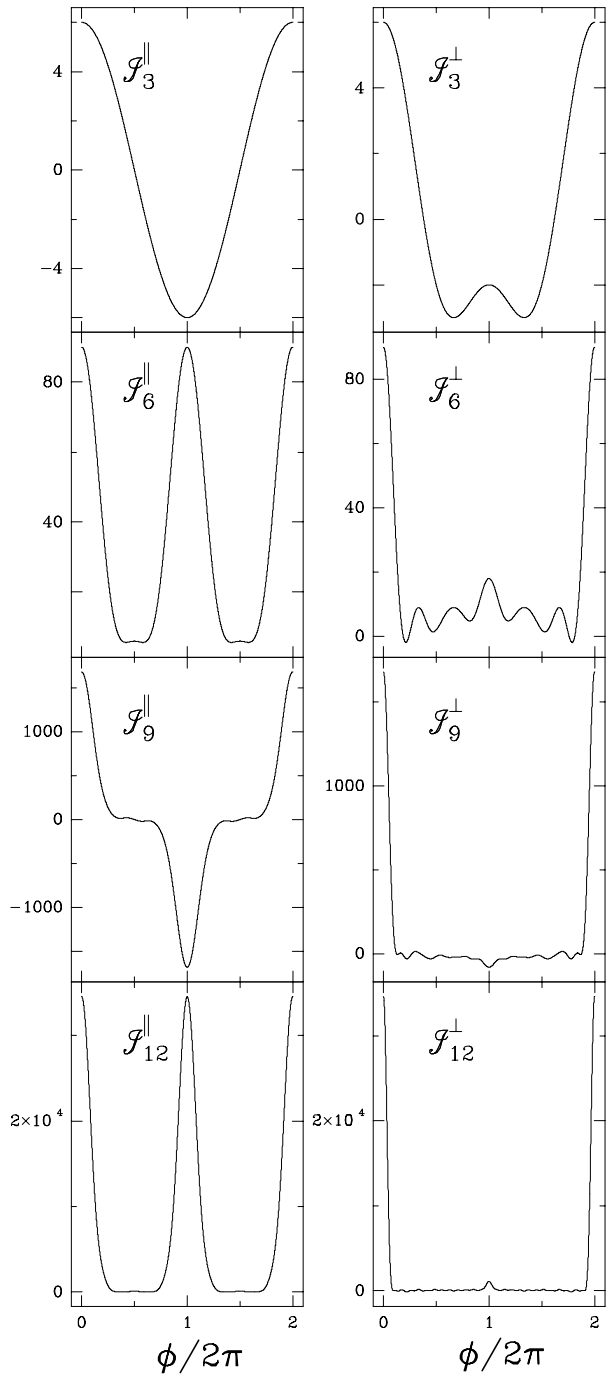
\begin{figure}
\caption{Sums over forward-scattering paths between two 
diagonally-separated sites on a 3D cubic lattice: ${\cal I}_{3}^{\parallel}$ 
through ${\cal I}_{12}^{\parallel}$ for ${\bf B}=(\phi,\phi,\phi)$ and 
${\cal I}_{3}^{\perp}$ through ${\cal I}_{12}^{\perp}$ for 
${\bf B}=(\phi/2,\phi/2,-\phi)$, as functions of $\phi/2\pi$. 
Note that while ${\cal I}_{3m}^{\parallel}$ has the $2\pi$ ($4\pi$) 
periodicity for even (odd) $m$, ${\cal I}_{3m}^{\perp}$ always has a 
period $4\pi$.}
\label{fig7}
\end{figure}

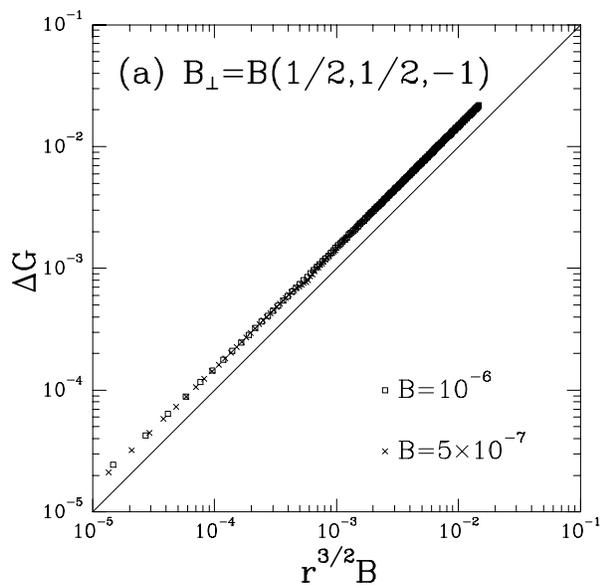
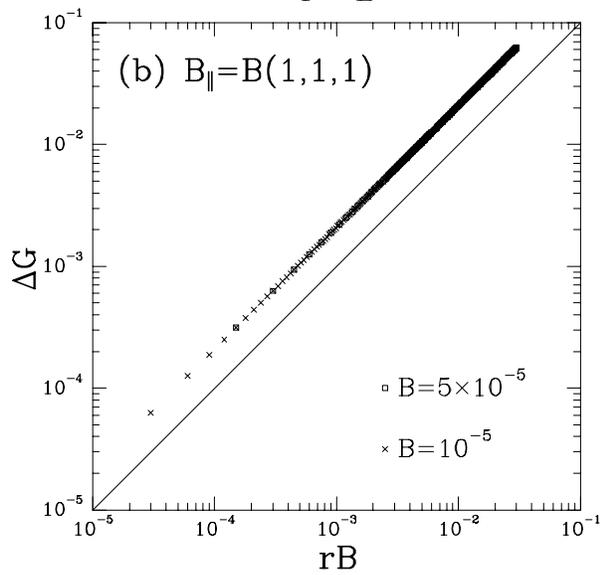
\begin{figure}
\caption{(a) $\Delta G$ versus $r^{3/2}B$ for 
${\bf B}_{\perp}=B\,(1/2,1/2,-1)$, and (b) $\Delta G$ versus $rB$ 
for ${\bf B}_{\parallel}=B\,(1,1,1)$ for several values of $B$ and hopping 
length $r=3, 6,\ldots, 600$. The collapse of all the data into a straight 
line verifies the scaling of small-field $\Delta G: 
(\protect\sqrt{2}/6)r^{3/2}\phi$ for ${\bf B}_{\perp}$  and 
$(1/3)r\phi$ for ${\bf B}_{\parallel}$. The distance between these data 
and the solid reference line reflects the prefactor: 
$\protect\sqrt{2}/6$ in (a) and $1/3$ in (b).}
\label{fig8}
\end{figure}

\end{document}